\UseRawInputEncoding
\documentclass[%
 twocolumn,
nofootinbib,
 amsfonts,amsmath,amssymb,
 aps,
prd,
floatfix,
showpacs,
longbibliography
]{revtex4-2}

\usepackage{graphicx}% Include figure files
\usepackage{dcolumn}% Align table columns on decimal point
\usepackage{bm}% bold math
\usepackage{wasysym}% Includes astro symbols
\usepackage{soul}% Highlighting
\usepackage{xfrac}
\usepackage[dvipsnames]{xcolor}
\definecolor{ReflexBlue}{rgb}{ .0902,.0902,.5882}
\usepackage[breaklinks]{hyperref}% add hypertext capabilities
\hypersetup{
    colorlinks=true,
    linkcolor=ReflexBlue,
    filecolor=magenta,      
    urlcolor=ReflexBlue,
    citecolor=ReflexBlue,
}
\usepackage{tikz}
\usetikzlibrary{decorations.pathmorphing}
\usepackage{mathrsfs}
\newcommand{\penrose}{%
\node (ext) at (\y,-\y) {};
\node (int) at (-\y,\y) {};
\path (int)+(90:2*\y)  coordinate (inttop)
           +(-90:2*\y) coordinate (intbot)
           +(0:2*\y)   coordinate (intright)
           +(180:2*\y) coordinate (intleft);
\draw (intleft)--(inttop)--(intright)--(intbot)--(intleft)--cycle;
\path (ext)+(90:2*\y)  coordinate (exttop)
           +(-90:2*\y) coordinate (extbot)
           +(180:2*\y) coordinate (extleft)
           +(0:2*\y)   coordinate (extright);
\draw (extleft)--(exttop)--(extright)--(extbot)--(extleft)--cycle;
}

\def\Tpext{\tikz[baseline=-0.4ex]{
\def\y{4pt}
\penrose
\path (ext)+(\y,\y)   coordinate (a)
           +(-\y,-\y) coordinate (b);
\draw[orange,thick] (a)--(b);
}}
\def\Rmext{\tikz[baseline=-0.4ex]{
\def\y{4pt}
\penrose
\path (ext)+(-\y,-\y) coordinate (a)
           +(0,0)     coordinate (b)
           +(-\y,\y)  coordinate (c);
\draw[orange,thick] (a)--(b)--(c);
}}
\def\Tpint{\tikz[baseline=-0.4ex]{
\def\y{4pt}
\penrose
\path (int)+(\y,\y)   coordinate (a)
           +(-\y,-\y) coordinate (b);
\draw[orange,thick] (a)--(b);
}}

\def\TmextRpint{\tikz[baseline=-0.4ex]{
\def\y{4pt}
\penrose
\path (int)+(3*\y,-3*\y)  coordinate (a)
           +(0,0)         coordinate (b)
           +(\y,\y)       coordinate (c);
\draw[orange,thick] (a)--(b)--(c);
}}

\def\Rmint{\tikz[baseline=-0.4ex]{
\def\y{4pt}
\penrose
\path (int)+(-\y,-\y) coordinate (a)
           +(0,0)     coordinate (b)
           +(-\y,\y)  coordinate (c);
\draw[orange,thick] (a)--(b)--(c);
}}
\def\RmextTmint{\tikz[baseline=-0.4ex]{
\def\y{4pt}
\penrose
\path (ext)+(-\y,-\y)    coordinate (a)
           +(0,0)        coordinate (b)
           +(-3*\y,3*\y) coordinate (c);
\draw[orange,thick] (a)--(b)--(c);
}}
\def\RmextRpint{\tikz[baseline=-0.4ex]{
\def\y{4pt}
\penrose
\path (ext)+(-\y,-\y)    coordinate (a)
           +(0,0)        coordinate (b)
           +(-2*\y,2*\y) coordinate (c)
           +(-\y,3*\y)   coordinate (d);
\draw[orange,thick] (a)--(b)--(c)--(d);
}}

\begin{document}

\title{Hawking radiation inside a charged black hole}

\author{Tyler McMaken}
 \email{Tyler.McMaken@colorado.edu}
 %\altaffiliation[Also at ]{Physics Department, XYZ University.}
\author{Andrew J. S. Hamilton}
 \email{Andrew.Hamilton@colorado.edu}
\affiliation{%
 JILA and Department of Physics, University of Colorado, Boulder, Colorado 80309, USA
}%
\date{\today}

\begin{abstract}
Here we analyze the Hawking radiation detected by an inertial observer in an arbitrary position in a Reissner-Nordstr\"om spacetime, with special emphasis on the asymptotic behavior of the Hawking spectrum as an observer approaches the inner or outer horizon. Two different methods are used to analyze the Hawking flux: first, we calculate an effective temperature quantifying the rate of exponential redshift experienced by an observer from an emitter's vacuum modes, which reproduces the Hawking effect provided the redshift is sufficiently adiabatic. Second, we compute the full Bogoliubov graybody spectrum observed in the three regimes where the wave equation can be solved analytically (at infinity and at the outer and inner horizons). We find that for an observer at the event horizon, the effective Hawking temperature is finite and becomes negative when $(Q/M)^2>8/9$, while at the inner horizon, the effective temperature is always negative and infinite in every direction the observer looks, coinciding with an ultraviolet-divergent spectrum.
\end{abstract}

%\keywords{Suggested keywords}

\maketitle

\section{\label{sec:int}Introduction}

Some of the most extraordinary effects in the study of quantum field theory in curved spacetime occur near the horizons of black holes. A black hole's event horizon is well known to exhibit a characteristic peeling of null geodesics that results in the detection of field radiation asymptotically far from the black hole \cite{haw74,haw75}. This radiation, known as Hawking radiation, has a thermal distribution (in the geometric optics limit), with a temperature directly proportional to the black hole's surface gravity at the event horizon.

Despite the success and relative robustness of Hawking's calculation, much debate has continued to this day concerning the nature, origin, and implications of Hawking radiation. From the perspective of quantum information theory, a driving question has been to understand how black holes evolve unitarily in spite of their seemingly thermal, informationless radiation. Though the calculations given in this paper give no quantitative measure of entropy and thus cannot address this problem directly, it may be that the increasing (and eventually diverging) presence of Hawking radiation we find as one ventures farther into a black hole's interior is closely tied to the mediation of unitary evolution, or at the least helps explain the vast number of degrees of freedom a black hole is expected to host.

More relevant to the present study, if the steady Hawking flux at late times is to be taken literally as originating from the event horizon, one might expect a local infaller at that horizon to detect a diverging number of particles seeping out with an initially divergent blueshift. However, subsequent derivations of Hawking radiation that rely more heavily on local phenomena, such as particles tunneling across the horizon \cite{par00}, pairs ripped apart by gravitational tidal forces \cite{dey17}, and particles created locally via the renormalized stress-energy tensor \cite{gid16}, find no divergence at the event horizon, but rather a modest ``quantum atmosphere'' of Hawking particles produced in its vicinity.

Though Hawking's original calculation only applies for radiation seen asymptotically far from a black hole, a formalism to study the Hawking radiation detected by an arbitrary observer any distance from a black hole was introduced in Refs.~\cite{bar11a,bar11b}. Those authors define an ``effective temperature'' function $\kappa$ that reduces to the surface gravity when it is sufficiently adiabatic over an interval (as will be detailed in Sec.~\ref{subsec:foreff}). This effective temperature is simply a measure of the rate of exponential redshifting seen by an observer from modes climbing out of collapsing matter's gravitational potential. In the context of quantum field theory in curved spacetime, such a redshifting, which need not rely on the actual formation of an event horizon, implies an excitation out of the vacuum state (i.e.\ particle production) due to the mixing of positive and negative frequencies. For a Schwarzschild black hole, as a freely falling observer is taken to infinity, $\kappa$ approaches the surface gravity $1/(4M)$\footnote{Throughout this paper we use the ${({-}{+}{+}{+})}$ metric signature and geometric units where ${c=G=\hbar=k_B=1}$.} predicted by Hawking, and as an observer approaches the event horizon, $\kappa$ is coincidentally found to approach a value exactly four times the surface gravity \cite{bar11c}. One of the main goals of the present paper is to extend this formalism to a charged (Reissner-Nordstr\"om) black hole.

A key difference between a Schwarzschild and Reissner-Nordstr\"om black hole is the presence in the latter case of a Cauchy horizon, a null hypersurface in the black hole's interior beyond which an observer can see a timelike singularity. Such a horizon, which will subsequently be referred to as the ``inner horizon,'' is of considerable interest both because of its presence in the astrophysically relevant Kerr metric and because of the open problem related to its stability in the presence of perturbative classical or quantum effects.

In 1968, Penrose was the first to point out that the inner horizon is a surface of infinite blueshift \cite{pen68,sim73}. Any external perturbations to the spacetime will produce ingoing radiation that an outgoing observer approaching the inner horizon will detect with diverging energy. Poisson and Israel pioneered the first full nonlinear analysis of this instability in 1990 \cite{poi90,bar90}, giving it the name ``mass inflation'' in reference to the exponentially inflated Misner-Sharp mass parameter \cite{mis64} measured at the inner horizon. As a result, the spacetime geometry near the inner horizon will classically break down and collapse to form a strong, spacelike singularity \cite{gne93,bur02,bur03,ham10,ham11a,ham11b,ham11c,ham17,che20,mcm21}, potentially alongside a weak, null singularity at late times \cite{ori91,ori92,bra95,ori96,ori98,ori99,daf05}.

Because of the unstable nature of the inner horizon in classical models, studies of quantum effects at the inner horizon have been ongoing. Early studies modeled the quantum effect of pair creation from the black hole's electric field by replacing the near-inner-horizon regime with a Schwarzschild-type solution once the electric field exceeded a critical value \cite{nov80,her94}, and later numerical studies with dynamical evolution found that Schwinger pair creation does indeed cause the inner horizon to form a spacelike singularity, but with a weak, null portion still surviving depending on the critical value of the electric field \cite{sor01a}. Those same authors also pioneered the first numerical study of Hawking radiation at the inner horizon \cite{sor01b}, with the result that a spacelike surface with diverging (super-Planckian) curvature forms \cite{hon10}.

However, these studies of Hawking radiation relied on the use of the 1+1D renormalized stress-energy tensor ${\langle T_{\mu\nu}\rangle}$ of a quantized scalar field to estimate the semiclassical backreaction. Renormalization of the full 3+1D stress-energy tensor for a black hole spacetime is a difficult problem with no known analytic solution; only recently have numerical studies begun to calculate the quantum fluxes of ${\langle T_{uu}\rangle}$ and ${\langle T_{vv}\rangle}$ at the inner horizon, generically finding divergences \cite{zil20,hol20}. Of particular note for the present study is the finding in Ref.~\cite{zil20} that the flux components of ${\langle T_{\mu\nu}\rangle}$ in double-null coordinates become negative for sufficiently large charge-to-mass ratios (${Q/M\gtrsim0.97}$). When taking into account first-order backreaction effects, this negative stress-energy implies the local abrupt expansion of the inner horizon geometry (see also \cite{lan19,tay20,bar21,arr21a,arr21b,zil21,kle21}).

Instead of focusing on the quantum renormalized stress-energy tensor, we here study the particle perception effects of Hawking radiation that do not rely on the ambiguities of renormalization in curved spacetime. Our choice of formalism also allows for a straightforward extension to the Hawking radiation seen in an arbitrary viewing direction, so that we may answer the question of whether the radial modes assumed in virtually all analyses of Hawking radiation are actually the dominant source of feedback at the inner horizon (especially since, for an outgoing observer across the inner horizon, only an exponentially small portion of the field of view is taken up by the blueshifting sky radially overhead). Following up on the study of the Schwarzschild interior in Ref.~\cite{ham18}, we extend those results to the Reissner-Nordstr\"om interior and focus on the seemingly paradoxical result that the effective Hawking temperature seen by an inertial observer always becomes negative and diverges at the inner horizon when the black hole has nonzero charge.

The conclusion that Hawking radiation diverges and possesses a negative temperature at the inner horizon of a Reissner-Nordstr\"om black hole highlights the need for a more realistic, dynamical model to describe the singular behavior of an astrophysical black hole near its inner horizon. The present study seeks to analyze this singular behavior in order to learn more about the constraints semiclassical physics imposes on near-inner horizon geometries.

Before diving into the bulk of the paper, it is worth pausing to comment on the implications (and especially the nonimplications) of a negative Hawking temperature. Hawking radiation is often pictured as a positive flux of particles escaping a black hole's horizon, coinciding with a negative flux of partner particles traveling inward to the black hole's singularity \cite{haw75}. However, the negative-temperature Hawking flux analyzed here is not simply an observation of the inward-traveling negative-energy Hawking partners. In contrast, our negative temperature will be found in both the ingoing and outgoing radiation sectors, and further, our calculations do not involve any tunneling across horizons. It may still be possible to formulate a local picture for the global calculations done here, but instead of the simple pair splitting at the outer horizon, one should imagine that virtual particle pairs created anywhere near and inside the black hole will be perturbed by radial gravitational tidal forces, and a negative temperature is realized because these forces will begin compressing instead of stretching once an observer comes close enough to the inner horizon \cite{cri16,ong20}.

How then should one interpret a negative Hawking temperature under the present formalism? The most straightforward answer is that the modes reaching an observer are blueshifting instead of redshifting, and this blueshift will result in a change in sign of the effective temperature of Eq.~(\ref{eq:kappa_tau}) below. However, the thermodynamic implications of such a change in sign are less apparent. Ref.~\cite{cur79} was the first to comment on the implications of the fact that the surface gravity $\varkappa_-$ defined at the inner horizon is negative, and many authors since have attempted to provide a consistent thermodynamic picture of a black hole with a negative-temperature inner horizon \cite{cve18,pel06,par08,tia17}. However, here we make no claims based on the Bekenstein-Hawking entropy nor any thermodynamic laws, and we also do not rely on any assumptions about what happens beyond the inner horizon. It may well be that the negative surface gravity has some implication for the temperature of a purely mathematical, analytically extended white hole emerging from an inner horizon. Nonetheless, the inner horizon effective temperature $\kappa$ describing the experience of an infalling observer is distinct from the global surface gravity $\varkappa_-$, and in fact $\kappa$ will be found either to diverge at the inner horizon or to equal some constant multiple of $\varkappa_-$ (see Sec.~\ref{subsubsec:kappa_rm}), depending on whether the observer looks up or down.

The structure of this paper is as follows: we begin in Sec.~\ref{sec:for} with a preliminary discussion of the effective temperature formalism used to calculate the Hawking radiation, then we proceed to calculate the effective temperature for various charges and observer positions in Sec.~\ref{sec:redshift}, commenting on validity of the adiabatic approximation in Sec.~\ref{subsec:adi} and generalizing from radial modes to arbitrary viewing directions in Sec.~\ref{subsec:ang}. Finally, in Sec.~\ref{sec:spe} we extend beyond the geometric optics approximation to calculate the full Bogoliubov spectrum in the asymptotic regimes where the scattering modes become simple (namely at infinity, the event horizon, and the inner horizon), and we conclude with a discussion in Sec.~\ref{sec:dis}.

\section{\label{sec:for}Formalism}

\subsection{\label{subsec:foreff}Defining an effective temperature as the rate of exponential redshift}

The Hawking flux perceived by a timelike geodesic observer in a black hole spacetime can be calculated through the use of an effective temperature function
\begin{equation}\label{eq:kappa_u}
    \kappa(u)\equiv-\frac{d}{du}\ln \left(\frac{dU}{du}\right)%=-\frac{d^2U}{du^2}\left(\frac{dU}{du}\right)^{-1}
\end{equation}
where the outgoing null coordinate $u$ gives the observer's position and the null coordinate $U$ gives the position of an emitter that defines the vacuum state \cite{bar11a,bar11b}. By a slight abuse of notation, the two worldlines labeled by coordinates $U$ and $u$ are connected by a null ray encoded by the function $U(u)$, and as long as $\kappa(u)$ remains approximately constant over a small interval around some point $u_*$, it directly implies that the vacuum expectation value of the particle number operator is consistent with that of a Planckian spectrum with temperature
\begin{equation}
    T_H(u_*)=\frac{\kappa(u_*)}{2\pi}.
\end{equation}
The constancy condition can be quantified by the adiabatic control function
\begin{equation}
    \epsilon(u)\equiv\frac{1}{\kappa^2}\left|\frac{d\kappa}{du}\right|,
\end{equation}
which must satisfy $\epsilon(u_*)\ll1$ in order for a thermal Hawking flux to be detected at $u_*$ \cite{bar11c}. However, even if the adiabatic condition is not satisfied, a nonzero $\kappa$ still implies the detection of particles corresponding to a nonzero Bogoliubov coefficient $\beta$; the only difference is that the spectral content will generally be non-Planckian.

Since both the observer and emitter can naturally use their proper times $\tau_\text{ob}$ and $\tau_\text{em}$ to label the different null rays they encounter throughout their journey, Eq.~(\ref{eq:kappa_u}) can be recast in a more intuitive form:
\begin{equation}\label{eq:kappa_tau}
    \kappa=-\frac{d}{d\tau_\text{ob}}\ln\left(\frac{\omega_\text{ob}}{\omega_\text{em}}\right),
\end{equation}
where the frequency $\omega$ (with either subscripts ``ob'' or ``em,'' which will be dropped hereafter when either label could apply), defined by
\begin{equation}
    \omega\equiv-k^\mu\dot{x}_\mu,
\end{equation}
is the temporal component of a null particle's coordinate 4-velocity ${k^\mu\equiv dx^\mu/d\lambda}$, measured in the frame of an observer or emitter with coordinate 4-velocity ${\dot{x}^\mu\equiv dx^\mu/d\tau}$. Eq.~(\ref{eq:kappa_tau}) makes it apparent that the effective temperature $\kappa$ is nothing more than a measure of the rate of frequency redshifting seen by an observer, an indicator of the exponential peeling of null rays first noted by Hawking as the crucial feature of black hole horizons responsible for particle creation \cite{haw74,haw75}.

For black hole spacetimes with a Killing horizon, in the limit as an observer approaches future timelike infinity, the notion of the effective temperature $\kappa(\tau)$ defined above coincides precisely with the notion of the surface gravity $\varkappa$ used to define a black hole's Hawking temperature \cite{bar11a}. Thus, $\kappa(\tau)$ provides a generalization of the Hawking effect for arbitrary observers around or inside of a black hole.

\subsection{\label{subsec:vac}Vacuum states}

Instead of performing calculations in a fully dynamical collapse spacetime, it is common to formulate an equivalent problem in an empty, eternal black hole spacetime like the Schwarzschild metric \cite{unr76}. As a result, the collapsing body must be replaced by appropriate boundary conditions on the past horizon, and these boundary conditions define the quantum field's vacuum state in that spacetime. Three options are generally discussed in the literature: the Boulware state, in which the quantum field's modes are defined to be positive frequency with respect to the Killing vector $\partial/\partial t$ on both the past horizon and past null infinity; the Hartle-Hawking state, in which modes are defined to be positive frequency with respect to the past boundaries' canonical affine coordinates\footnote{For example, for a Schwarzschild black hole, ${U=-4M\text{e}^{-u/(4M)}}$ is the outgoing Kruskal-Szekeres coordinate, whose vector field ${\partial/\partial U}$ is of Killing type on the past horizon. Positive frequency modes are then defined to be the eigenfunctions of the Lie derivative of the field in the ${\partial/\partial U}$ direction.} ${\partial/\partial U}$ and ${\partial/\partial V}$; and the (past) Unruh state \cite{unr76}, in which modes are defined to be positive frequency with respect to ${\partial/\partial U}$ on the past horizon and ${\partial/\partial t}$ at past null infinity. The last of these states is the one that is most physically relevant to the production of a Hawking flux to the future of a collapsing black hole and is the state that will be employed here.

In the effective temperature framework, the vacuum state is specified by the spacetime position and state of motion (the orbital parameters) of the emitter. For example, the Boulware state corresponds to a static emitter maintaining a constant radius $r_0$. This state is thus only defined for the exterior portion of the black hole, since an emitter cannot remain static below the event horizon. A freely falling observer measuring in the Boulware state will see diverging stress-energy at the horizon, as a result of the diverging acceleration required for the Boulware emitter to remain static there.

In contrast, the Unruh state is associated with a freely falling emitter, positioned either at the black hole's horizon or at infinity. The outgoing Unruh modes correspond to the limit ${r_\text{em}\to r_+}$, so that the observer sees the emitter frozen on the past horizon (one may equivalently take the Unruh emitter's descent into the black hole to have occurred sufficiently far into the past), and the ingoing Unruh modes correspond to the limit ${r_\text{em}\to\infty}$, so that the observer sees the emitter safely resting in the sky above. Since the observer and the Unruh emitter are generally not located at the same spacetime coordinate (as in the Boulware state), their modes must be connected via a null geodesic, since the quantum field under study here is massless.

To see how the choice of vacuum corresponds to the specification of the emitter's worldline, consider an emitter radially free-falling from rest at infinity\footnote{The same arguments should hold for any inertial free-faller; here we present the radial, ${E=1}$ case for simplicity.} into a static, asymptotically flat, spherically symmetric black hole, which is given by the line element 
\begin{equation}\label{eq:line_element_RN}
    ds^2=-\Delta(r)\ dt^2+\frac{dr^2}{\Delta(r)}+r^2\left(d\theta^2+\sin^2\!\theta d\phi^2\right).
\end{equation}
The horizon function ${\Delta(r)}$ has the property that it vanishes linearly as $r$ approaches a horizon, and it asymptotes to unity as ${r\to\infty}$.

Such an emitter will have coordinate 4-velocity with nonzero components
\begin{subequations}\label{eq:RN_4-velocity}
\begin{gather}
    \dot{t}\equiv\frac{dt}{d\tau}=\frac{1}{\Delta},\displaybreak[0]\\
    \dot{r}\equiv\frac{dr}{d\tau}=-\sqrt{1-\Delta}.
\end{gather}
\end{subequations}

When the emitter is at infinity (${\Delta\to1}$) sending modes inward, Eq.~(\ref{eq:RN_4-velocity}a) implies that the emitter's proper time $\tau$ will tick at the same proportionate rate as the global timelike Killing coordinate $t$. Thus, $t$ will be the coordinate the emitter uses to define positive frequency, just as expected for ingoing Hawking modes originating from past null infinity.

However, when the emitter reaches a horizon (${\Delta\to0}$), Eq.~(\ref{eq:RN_4-velocity}a) implies that the static Schwarzschild time $t$ will tick at an infinitely faster rate than the emitter's proper time $\tau$. So heuristically, instead of seeing wave modes of the form ${\exp(-i\omega t)}$, the emitter should end up seeing modes of the form ${\exp[-i\omega\exp(-kt)]}$ (for some constant $k$), so that even when $t$ diverges, the emitter's proper time will still remain finite. The new time coordinate defined by these modes will be found to coincide with the oft-studied Kruskal-Szekeres coordinate $U$.

To make the above arguments more precise, and to extend the discussion to distinguish ingoing and outgoing modes (which depend on both the emitter's proper time and the proper distance between wavefronts), consider a set of eikonal waves in the emitter's locally orthonormal tetrad frame $\{\bm{\gamma}_0,\bm{\gamma}_1,\bm{\gamma}_2,\bm{\gamma}_3\}$, whose tangent-space coordinates will be labeled $\xi^0$, $\xi^1$, $\xi^2$, and $\xi^3$. This tetrad frame is constructed so that it is continuous across the event horizon and so that the time axis $\bm{\gamma}_0$ is always timelike and future-directed, while the radial axis $\bm{\gamma}_1$ is always spacelike and outward-directed. In the limit of large frequency $\omega$, to leading order in ${1/\omega}$, the ingoing ($+$) or outgoing ($-$) components of the eikonal wavefront will follow a null geodesic congruence with tetrad-frame 4-momentum (neglecting any normalization factors)
\begin{equation}\label{eq:km}
    k^{\hat{m}}\equiv\frac{d\xi^{\hat{m}}}{d\lambda}=\left(1,\ \pm1,\ 0,\ 0\right).
\end{equation}
The transformation from the emitter's local tetrad frame to a coordinate frame can be accomplished through the use of the appropriate vierbein. For an external\footnote{The case of a free-faller in the black hole interior follows the same line of reasoning as the exterior case presented here, \textit{mutatis mutandis}.} radial free-faller with specific energy $E$ (where ${E=1}$ corresponds to rest at infinity), in the static polar spherical chart this vierbein reads
\begin{equation}
    e^{\hat{m}}_{\ \ \mu}=
    \begin{pmatrix}
        E & \sqrt{E^2-\Delta} & 0 & 0\\
        \sqrt{E^2-\Delta} & E & 0 & 0\\
        0 & 0 & r & 0\\
        0 & 0 & 0 & r\sin\theta
    \end{pmatrix}
\end{equation}
(where rows label the coordinates $\xi^0$, $\xi^1$, $\xi^2$, $\xi^3$ of the emitter's locally inertial frame, and columns label the global coordinates $t$, $r^*$, $\theta$, $\varphi$). Here we define the tortoise coordinate $r^*$ by
\begin{equation}\label{eq:tortoise}
    \frac{dr}{dr^*}=\Delta.
\end{equation}
The coordinate-frame 4-momentum ${k^\mu=k^{\hat{m}}e_{\hat{m}}^{\ \ \mu}}$ then follows immediately:
\begin{equation}\label{eq:kmu}
    k^\mu=\left(\frac{E\mp\sqrt{E^2-\Delta}}{\Delta},\ \pm\frac{E\mp\sqrt{E^2-\Delta}}{\Delta},\ 0,\ 0\right).
\end{equation}
If the emitter defines some positive frequency $\omega$ (along with the corresponding wave number $\omega/c$), then their natural choice of ingoing (upper sign) or outgoing (lower sign) modes will take the form ${\exp[-i\omega(\xi^0\pm\xi^1)]}$, which can be written in coordinate form by matching the affine distances of Eqs.~(\ref{eq:km}) and (\ref{eq:kmu}):
\begin{equation}\label{eq:tetrad1forms}
    d\xi^0\pm d\xi^1=\frac{\Delta}{E\mp\sqrt{E^2-\Delta}}\left(dt\pm dr^*\right).
\end{equation}
Asymptotically, as a unit-energy emitter approaches infinity (${\Delta\to1}$), the fraction in Eq.~(\ref{eq:tetrad1forms}) reduces to unity, so that the proper choice of coordinates to define Unruh modes at infinity is the Eddington-Finkelstein double null coordinate system, defined in both the exterior and interior as
\begin{equation}
    u\equiv t-r^*,\quad v\equiv t+r^*,
\end{equation}
where $u$ here is the same outgoing null coordinate as in Eq.~(\ref{eq:kappa_u}).

When the emitter is at a horizon (${\Delta\to0}$), the mode behavior depends on whether the waves are ingoing or outgoing. For the ingoing modes of a positive-energy free-faller or the outgoing modes of a negative-energy free-faller (neither of which are needed to define an Unruh emitter but will prove useful later to define the natural modes seen by horizon observers), the fraction in Eq.~(\ref{eq:tetrad1forms}) reduces to $2E$, so that the proper modes (after $\omega$ is properly scaled) are once again the Eddington-Finkelstein modes ${\exp[-i\omega(t\pm r^*)]}$.

But for the outgoing modes of a positive-energy free-faller or the ingoing modes of a negative-energy free-faller at the horizon, the fraction in Eq.~(\ref{eq:tetrad1forms}) vanishes, so a more appropriate coordinate choice must be found. Define a new coordinate $\bar{U}$ such that the outgoing Unruh modes at the horizon will be written as ${\exp[-i\omega\bar{U}]}$. Then Eq.~(\ref{eq:tetrad1forms}) implies that $\bar{U}$ must satisfy
\begin{equation}\label{eq:dU/du}
    \frac{d\bar{U}}{du}\underset{\Delta\to0}{=}\frac{\Delta}{2E}\approx\frac{r-r_\pm}{2E}\frac{d\Delta}{dr}\bigg|_{r_\pm}
\end{equation}
in the near-horizon limit. From this expression one can identify the quantity 
\begin{equation}\label{eq:surface_gravity}
    \varkappa_\pm\equiv\frac12\frac{d\Delta}{dr}\bigg|_{r_\pm}
\end{equation}
as the outer ($+$) or inner ($-$) horizon's surface gravity. For an emitter with ${E=1}$, since from Eqs.~(\ref{eq:km}), (\ref{eq:tortoise}), and (\ref{eq:kmu}), the radius $r$ is related to the horizon-limit outgoing proper null coordinate $\bar{U}$ by
\begin{equation}
    \frac{dr}{d\bar{U}}=-\frac{1+\sqrt{1-\Delta}}{2}\underset{\Delta\to0}{=}-1,
\end{equation}
then Eq.~(\ref{eq:dU/du}) solves as
\begin{equation}\label{eq:U(u)}
    \bar{U}\propto\exp\left(-\varkappa_\pm u\right).
\end{equation}
Eq.~(\ref{eq:U(u)}) assumes that $\bar{U}$ is chosen to begin at 0 at the event horizon, when ${u\to\infty}$. This form of the emitter's proper time (up to an irrelevant normalization factor) is precisely the form of the outgoing Kruskal-Szekeres coordinate $U$ used by Unruh to define positive frequency on the past horizon \cite{unr76}. Thus, the outgoing modes of the Unruh state correspond to those seen as positive frequency by an emitter in free fall asymptotically close to the past horizon.

In some sense, we have done nothing more than ``rederive the obvious'' in showing how one may obtain past Unruh null boundary conditions. However, in addition to providing yet another way of understanding the validity of this choice of vacuum state, the generalized derivation above also provides a natural specification of ingoing and outgoing modes for freely falling observers at either horizon, without any reliance on global Killing vector fields or asymptotically Minkowski regimes. We will return to this idea when solving the wave equation in Sec.~\ref{sec:spe}.

As a final comment concerning the choice of vacuum state, an additional family of vacuum states was used by Ref.~\cite{bar11c} to mimic the switching on of Hawking radiation as a black hole first forms during a collapse. These ``collapse vacua'' correspond to emitters in free fall from rest at infinity, each separated from the observer by a time delay $\delta\tau$ (as in the Unruh state), but not necessarily in the limit as they approach the horizon or infinity. However, in the present work, we are not concerned with the initial transient collapse dynamics of a black hole; rather, we will focus on the late-time steady-state behavior once the black hole has settled down into the Unruh state, which should occur only a few light-crossing times after the black hole's formation.

\section{\label{sec:redshift}Redshifting perceived by an infaller}

Here we examine the effective temperature seen by a freely falling inertial observer in a charged black hole spacetime with a quantized scalar field. In Sec.~\ref{subsec:rad} we calculate the effective temperature $\kappa$ for an observer looking in the radial direction via Eq.~(\ref{eq:kappa_tau}), in Sec.~\ref{subsec:adi} we analyze when this $\kappa$ satisfies adiabaticity, and in Sec.~\ref{subsec:ang} we generalize to an observer looking in an arbitrary direction.

\subsection{\label{subsec:rad}Radial effective temperature}

Consider the line element of Eq.~(\ref{eq:line_element_RN}), which describes the geometry of a charged, spherically symmetric black hole when the horizon function $\Delta$ takes the form
\begin{align}
    \Delta&=\left(1-\frac{r_+}{r}\right)\left(1-\frac{r_-}{r}\right),\\
    r_\pm&\equiv M\pm\sqrt{M^2-Q^2}.
\end{align}
The black hole modeled by this geometry is known as the Reissner-Nordstr\"om black hole, which possesses a mass $M$ and a charge $Q$. The length scales $r_+$ and $r_-$ are referred to respectively as the outer (event) horizon and the inner (Cauchy) horizon.

The rate of redshift seen by a radially infalling observer has already been calculated for the spacetime of Eq.~(\ref{eq:line_element_RN}) for arbitrary $\Delta$ (see Appendix B of Ref.~\cite{ham18}), though that analysis was only carried out explicitly for Schwarzschild (${Q/M=0}$). Here we quote the main results and specialize to Reissner-Nordstr\"om with a focus on the inner horizon.

The frequency $\omega$ measured in the frame of an observer ($\equiv\omega_\text{ob}$) or emitter ($\equiv\omega_\text{em}$) with specific energy $E$, normalized to the frequency $\omega_\infty$ seen at rest at infinity, is
\begin{equation}
    \frac{\omega}{\omega_\infty}=\frac{E\pm\sqrt{E^2-\Delta}}{\Delta},
\end{equation}
where the upper (lower) sign applies to outgoing (ingoing) null rays. The effective temperature $\kappa$ can then be calculated with the help of the chain rule:
\begin{align}
    \kappa&=-\frac{d}{d\tau_\text{ob}}\ln\left(\frac{\omega_\text{ob}}{\omega_\text{em}}\right)\nonumber\\
    &=-\omega_\text{ob}\left(\frac{\dot{r}_\text{ob}}{\omega_\text{ob}}\frac{\partial\ln\omega_\text{ob}}{\partial r_\text{ob}}-\frac{\dot{r}_\text{em}}{\omega_\text{em}}\frac{\partial\ln\omega_\text{em}}{\partial r_\text{em}}\right)\nonumber\\
    &=\mp\frac12\frac{\omega_\text{ob}}{\omega_\infty}\left(\frac{d\Delta_\text{ob}}{dr_\text{ob}}-\frac{d\Delta_\text{em}}{dr_\text{em}}\right),\label{eq:kappa_chainrule}
\end{align}
where an overdot signifies differentiation with respect to the observer's or emitter's proper time $\tau$.

For outgoing modes (upper sign), the Unruh emitter must be placed at the event horizon (${r_\text{em}\to r_+}$), and for ingoing modes (lower sign), the Unruh emitter resides at infinity (${r_\text{em}\to\infty}$). The result, for an observer in free fall from rest at infinity (${E_\text{ob}=1}$), is the sensation of two independent effective temperatures corresponding to the outgoing ($\kappa^+$) and ingoing ($\kappa^-$) Hawking modes (throughout the rest of this paper, $\pm$ superscripts will always refer to outgoing/ingoing quantities, while $\pm$ subscripts will always refer to outer/inner horizon quantities):
\begin{subequations}\label{eq:RN_kappa}
\begin{align}\label{eq:RN_kappa+}
    \kappa^+&=\frac{Mr_{\text{ob}}\left(1-r_{\text{ob}}^2/r_+^2\right)-Q^2\left(1-r_{\text{ob}}^3/r_+^3\right)}{r_{\text{ob}}^2\left(-r_{\text{ob}}+\sqrt{2Mr_{\text{ob}}-Q^2}\right)},\\\label{eq:RN_kappa-}
    \kappa^-&=\frac{Mr_{\text{ob}}-Q^2}{r_{\text{ob}}^2\left(r_{\text{ob}}+\sqrt{2Mr_{\text{ob}}-Q^2}\right)}.
\end{align}
\end{subequations}

The rest of this Sec.~\ref{subsec:rad} will be devoted to exploring the implications of Eqs.~(\ref{eq:RN_kappa}). As a first comment, because of the square root in the denominator, both temperatures become imaginary when the observer is located close enough to the origin, specifically when ${r_\text{ob}<Q^2/(2M)}$. However, such values of $r_\text{ob}$ are strictly less than the inner horizon radius $r_-$ for all choices of $Q$, and the failure of Eqs.~(\ref{eq:RN_kappa}) in this region coincides with the failure of Gullstrand-Painlev\'e coordinates in the same region, indicative of the presence of an unphysical negative interior mass $M(r)$ (i.e.\ this is where an infaller would bounce back due to the effects of the repulsive charged singularity on the spacetime) \cite{ham08}. Since the region below the inner horizon should be physically disregarded due to the semiclassical singular behavior examined below, that region will not be explored any further here.

\begin{figure*}[t]
\centering
\begin{minipage}[l]{0.85\columnwidth}
  \includegraphics[width=\columnwidth]{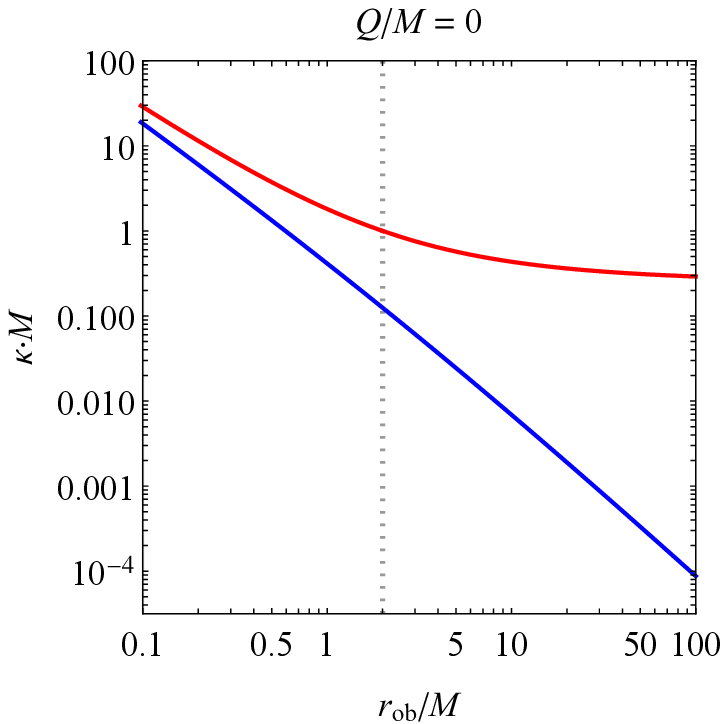}
\end{minipage}%
\hspace{1.5em}
\begin{minipage}[r]{0.85\columnwidth}
    \includegraphics[width=\columnwidth]{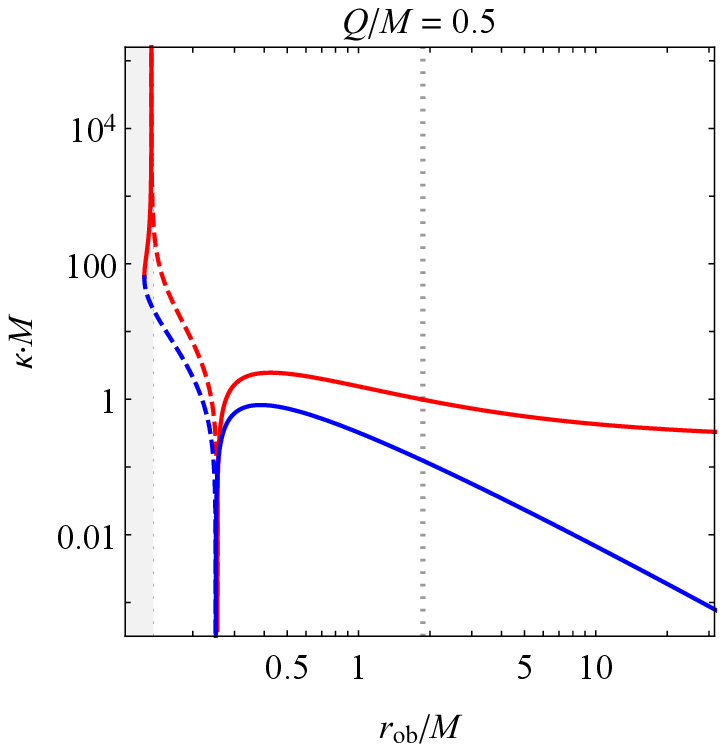}
\end{minipage}
\begin{minipage}[l]{0.85\columnwidth}
  \includegraphics[width=\columnwidth]{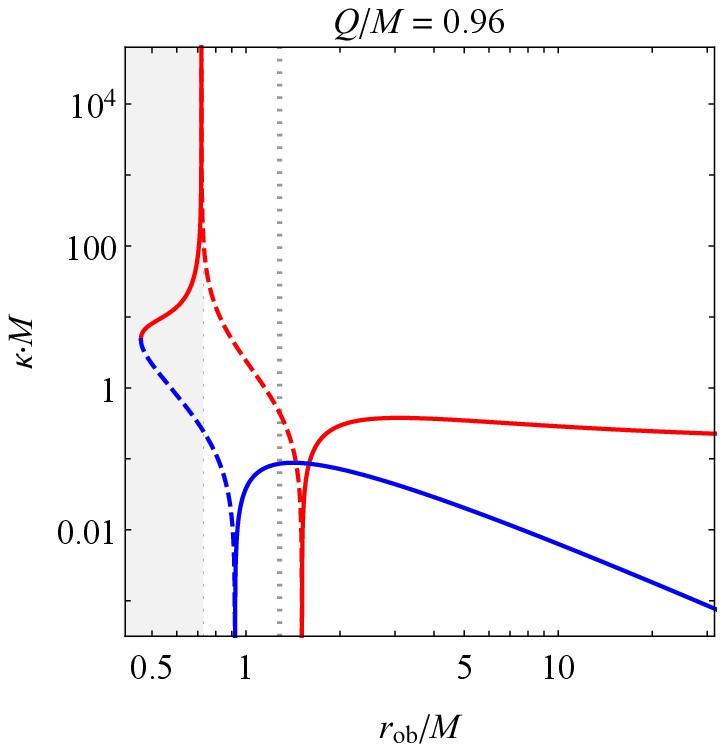}
\end{minipage}%
\hspace{1.5em}
\begin{minipage}[r]{0.85\columnwidth}
    \includegraphics[width=\columnwidth]{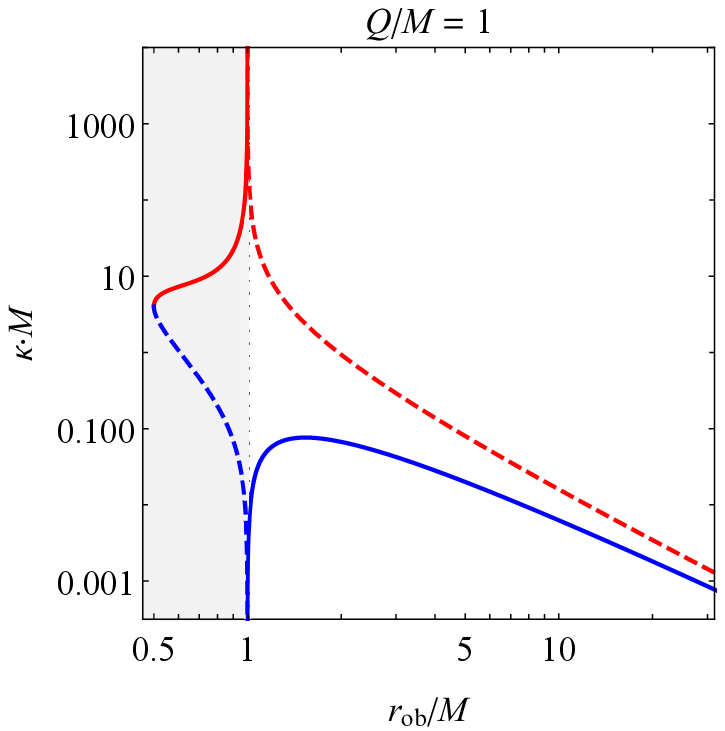}
\end{minipage}
\caption{Outgoing effective temperature $\kappa^+$ (red curve) and ingoing effective temperature $\kappa^-$ (blue curve) as a function of observer radius $r_\text{ob}$ for various choices of the Reissner-Nordstr\"om black hole charge $Q$, all in units of the black hole mass $M$. Solid curves indicate positive values on the log plot, and dashed curves indicate negative values. The inner and outer horizons are shown with gray, dotted vertical lines, and the unphysical region below the inner horizon is grayed out.\label{fig:kappa_RN}}
\end{figure*}

Second, it should be noted that for an observer asymptotically far from the black hole, the above formulas reproduce familiar results: the outgoing sector's temperature asymptotically approaches the black hole's surface gravity $\varkappa_+$ defined by Eq.~(\ref{eq:surface_gravity}), and the ingoing Hawking sector vanishes:
\begin{equation}\label{eq:RN_kappa_inf}
    \lim_{r_\text{ob}\to\infty}\left\{\kappa^+,\ \kappa^-\right\}=\left\{\frac{r_+-r_-}{2r_+^2},\ 0\right\}.
\end{equation}
As expected, $\kappa^+$ approaches ${1/(4M)}$ in the Schwarzschild ${Q/M=0}$ limit and vanishes in the extremal ${Q/M=1}$ limit. These limits can be seen in the respective panels of Fig.~\ref{fig:kappa_RN}, which shows the full behavior of ${\kappa^+(r_\text{ob})}$ and ${\kappa^-(r_\text{ob})}$ for different choices of the black hole's charge-to-mass ratio.

\subsubsection{Negative \texorpdfstring{$\kappa$}{kappa} at the event horizon and beyond}

\begin{figure}[t!]
  \includegraphics[width=0.93\columnwidth]{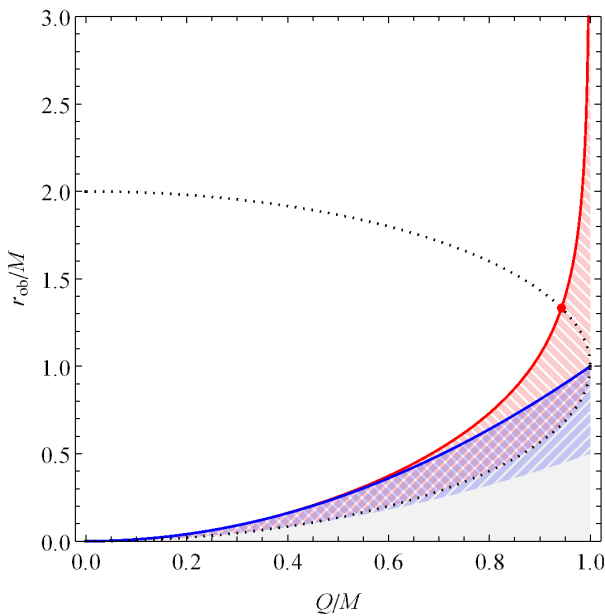}
  \caption{Regions of negative temperature in the Reissner-Nordstr\"om charge-radius parameter space. The black dotted curve shows the inner and outer horizons, which converge in the extremal limit ${Q/M=1}$. The red (blue) curve shows regions where the effective temperature in the outgoing sector $\kappa^+$ (ingoing sector $\kappa^-$) equals zero, and the red (blue) hatched shading shows regions where the effective temperature $\kappa^+$ ($\kappa^-$) is negative. The red dot marks the charge ${Q/M=\sqrt{8/9}}$ above which the effective temperature $\kappa^+$ becomes negative outside the event horizon. As in Fig.~\ref{fig:kappa_RN}, the unphysical region below the inner horizon is shaded out gray. \label{fig:kappa_RN_negregion}}
\end{figure}

As an observer freely falling from infinity approaches the Reissner-Nordstr\"om event horizon and enters the black hole, the effective Hawking temperatures $\kappa^+$ and $\kappa^-$ grow from their initial values at infinity until reaching a maximum value, after which they quickly drop to zero and become negative (excepting the special cases ${Q/M=0,1}$). When the observer crosses the event horizon, the effective temperatures in the outgoing and ingoing sectors are
\begin{equation}\label{eq:RN_kappa_r+}
    \lim_{r_\text{ob}\to r_+}\left\{\kappa^+,\ \kappa^-\right\}=\left\{\frac{2\left(r_+-2r_-\right)}{r_+(r_+-r_-)},\ \frac{r_+-r_-}{4r_+^2}\right\}.
\end{equation}
The most notable feature of Fig.~\ref{fig:kappa_RN} is the fact that $\kappa^+$ and $\kappa^-$ become negative (indicated by the dashed lines) if the observer is close enough to the inner horizon, corresponding to a blueshifting of the observed modes instead of the usual exponential redshifting. The exact regions with negative temperature depend heavily on the charge $Q$, generally extending farther outward with increasing charge. The ingoing radiation (the blue curve) has negative temperature only below the event horizon, coinciding exactly with the change in sign of the Weyl scalar at $r={Q^2/M}$, but curiously enough, the outgoing radiation (red) can have negative temperature even above the event horizon, and in the extremal case, the effective temperature in the entire exterior is negative. How large a charge is necessary for a negative temperature to be detected outside the black hole? From Eq.~(\ref{eq:RN_kappa_r+}), $\kappa^+$ will be negative above the event horizon if the event horizon is less than double the inner horizon's radius, which occurs when ${(Q/M)^2>8/9}$. This special value of $Q$ is shown in Fig.~\ref{fig:kappa_RN_negregion} with a red dot marking the intersection of the solid red and dotted black curves.

The value ${(Q/M)^2=8/9}$, where the event horizon coincides with the radial inflection point in the black hole's horizon function $\Delta$, has shown up previously in the literature for Reissner-Nordstr\"om black holes in varying contexts. Ong and Good \cite{ong20} used a heuristic gravitational analog of the Schwinger effect to show that the energy of two Hawking quanta split apart from tidal forces will be negative near the horizon when ${(Q/M)^2>8/9}$. This change in sign can be traced to the change in the radial tidal force, as measured by the proper acceleration of the free-fall-frame geodesic deviation vector, from the usual stretching force into a compressing force as $Q$ is increased \cite{cri16}. Similarly, the square of the free-fall temperature obtained by embedding the black hole in a six-dimensional flat spacetime and finding the Unruh temperature of the analogous observer was found to be negative for ${(Q/M)^2>8/9}$ \cite{bry08}, which those authors interpreted as a failure to detect any radiation. Finally, in the 1+1D case, the renormalized expectation values of the temporal and radial components of a scalar field's stress-energy tensor ${\langle T_\mu^{\ \nu}\rangle}$ become negative at the event horizon in the exact same range \cite{lor97}. These studies apply to a variety of different semiclassical effects that all point toward similar semiclassical behavior, but in the present case, the physical interpretation of a negative effective temperature $\kappa$ is not so clear-cut, especially given the lack of adiabaticity in some regions of interest (described in Sec.~\ref{subsec:adi}). A more robust physical interpretation is therefore deferred until the spectral analysis of Sec.~\ref{sec:spe}.

\subsubsection{Diverging \texorpdfstring{$\kappa$}{kappa} at the inner horizon\label{subsubsec:kappa_rm}}

Now, consider the effective temperature seen when the observer reaches the inner horizon. As can be seen from Figs.~\ref{fig:kappa_RN} and \ref{fig:kappa_RN_negregion}, both the outgoing and ingoing effective temperatures $\kappa^+$ and $\kappa^-$ are always nonpositive at the inner horizon.\footnote{Here we treat values of ${\kappa\to\pm\infty}$ as equivalent to maintain consistency with the standard entropic definition of temperature, where both coincide with zero inverse thermodynamic temperature $\beta$.} The effective temperature $\kappa^-$ for the ingoing sector remains finite for all nonzero values of $Q$, but the outgoing temperature $\kappa^+$ always diverges at the inner horizon. Defining a new coordinate ${z_\text{ob}\equiv(r_\text{ob}-r_-)/(r_+-r_-)}$ representing the observer's dimensionless distance above the inner horizon, in the limit of small ${z_\text{ob}\ll1}$, one has (to leading order in $z_\text{ob}$):
\begin{equation}\label{eq:RN_kappa_r-}
    \lim_{\overset{z_\text{ob}\to0,}{E_\text{ob}\to1}}\left\{\kappa^+,\ \kappa^-\right\}=\left\{-\frac{r_+^2+r_-^2}{r_+^2(r_+-r_-)z_\text{ob}},\ -\frac{r_+-r_-}{4r_-^2}\right\}.
\end{equation}
From Eq.~(\ref{eq:RN_kappa_r-}), one can see that the perceived temperature from outgoing radiation at the inner horizon (when the observer looks straight down at the past horizon) quickly approaches negative infinity, while the practically irrelevant perceived temperature from ingoing radiation (when the observer looks up at the sky above) equals half the inner horizon's surface gravity $\varkappa_-$ of Eq.~(\ref{eq:surface_gravity}).

Note that the above analysis applies only to an ingoing observer, who must pass through the left leg of the inner horizon (labeled $\cal{H}^-_{\textit{r}_-}$ in Fig.~\ref{fig:RN_penrose}). In order to reach the right leg of the inner horizon, an infalling observer must accelerate outward until they acquire negative energy as measured by another observer at infinity. For an observer with specific energy ${E_\text{ob}=-1}$ (who can exist only inside the event horizon, where the Killing time coordinate $t$ is spacelike), the only change to Eqs.~(\ref{eq:RN_kappa}) that is needed is to swap their denominators. With this change, the resulting effective temperatures for an outgoing observer at the inner horizon are
\begin{align}\label{eq:RN_kappa_r-_E-1}
    &\lim_{\overset{z_\text{ob}\to0,}{E_\text{ob}\to-1}}\left\{\kappa^+,\ \kappa^-\right\}=\nonumber\\
    &\left\{-\frac{r_+-r_-}{4r_-^2}\left(1-\frac{r_+-r_-}{r_+^2}\right),\ -\frac{1}{(r_+-r_-)z_\text{ob}}\right\}.
\end{align}

Both effective temperatures are still negative. The main change to be noticed when traveling through the right portion of the inner horizon instead of the left portion is that the ingoing effective temperature $\kappa^-$ seen from the sky above diverges instead of the outgoing temperature seen from the past horizon below. This divergence of $\kappa^-$ is consistent with the inner horizon blueshift divergence first noted by Penrose \cite{pen68}. In contrast, the outgoing effective temperature $\kappa^+$ remains finite for large $Q$, vanishing as ${Q/M\to1}$, though as ${Q/M\to0}$, $\kappa^+$ diverges (just as $\kappa^-$ does in the case of an ingoing observer at the inner horizon).

One final special case is an observer with zero energy, who passes through the intersection of the left and right legs of the inner horizon (the uppermost point in Fig.~\ref{fig:RN_penrose}). At this special location, the ingoing and outgoing effective temperatures both diverge:
\begin{equation}
    \lim_{\overset{z_\text{ob}\to0,}{E_\text{ob}\to0}}\left\{\kappa^+,\ \kappa^-\right\}=\left\{-\frac{r_+^2+r_-^2}{2r_+^2r_-\sqrt{z_\text{ob}}},\ -\frac{1}{2r_-\sqrt{z_\text{ob}}}\right\}.
\end{equation}
Thus, no matter what portion of the inner horizon the observer reaches, at least one of the Hawking sectors will always feature a divergent, negative temperature.

Divergent semiclassical behavior at the Reissner-Nordstr\"om inner horizon is already well anticipated in the literature. As early as 1980, it was argued that the renormalized expectation value of the stress-energy tensor in regular coordinates must diverge on at least one of the two legs of the inner horizon \cite{his80}. More recently, the renormalized stress-energy tensor in the Unruh state was computed explicitly at the inner horizon, and it was found generically to diverge \cite{zil20}. There are a few differences between that study's results and the results found here; namely, the sign of ${\langle T_{uu}\rangle^U_\text{ren}}$ and ${\langle T_{vv}\rangle^U_\text{ren}}$ at the inner horizon can be either positive or negative depending on the charge $Q$ (as opposed to the purely negative $\kappa^\pm$ found here), and those stress-energy tensor fluxes both vanish in the extremal limit (while only $\kappa^+$ vanishes as ${Q/M\to1}$ for outgoing observers) \cite{zil21}. However, the effective temperature and the renormalized stress-energy tensor should not be expected to agree, since the former describes the perception by an infaller of a spectral distribution while the latter describes the tensorial flux and energy density of that radiation\textemdash a perceptual formulation of ${\langle T_{\mu\nu}\rangle}$ would depend not only on $\kappa$ but also on $\dot{\kappa}$ \cite{bar16}.

\subsubsection{Dependence of \texorpdfstring{$\kappa$}{kappa} on the observer's energy}\label{subsubsec:kappa_E}

Finally, consider how the effective temperatures $\kappa^\pm$ given by Eq.~(\ref{eq:kappa_chainrule}) change for arbitrary observer energies. Can an observer eliminate the detection of Hawking radiation, or perhaps even change its sign, simply by Lorentz-boosting to a different frame?

The only contribution to the effective temperatures of Eq.~(\ref{eq:kappa_chainrule}) that depends on the observer's specific energy $E_\text{ob}$ is the factor $\omega_\text{ob}$, the observer-frame frequency. Thus, any Lorentz-boosting effects on the effective temperature seen by a radial observer are solely confined to those caused by a Doppler factor shift. This shift will never change the sign of $\kappa^\pm$ for an observer at a given radius; it will only change the overall magnitude. In particular, as the observer speeds up, in the limit ${E_\text{ob}\gg1}$ (or ${E_\text{ob}\ll-1}$), the magnitude of $\kappa^+$ (or $\kappa^-$, respectively) will increase linearly with $E_\text{ob}$. Similarly, in the limit ${E_\text{ob}\ll-1}$ (or ${E_\text{ob}\gg1}$), the magnitude of $\kappa^+$ (or $\kappa^-$, respectively) will drop reciprocally to zero. Between these two limits, $\kappa^\pm$ varies monotonically with $E_\text{ob}$, so even if an interior observer's energy passes through zero, $\kappa^\pm$ will always remain the same sign.

The change in sign in the radial effective temperature for an inertial observer is thus purely geometrical in origin. As an observer changes their energy (or even their viewing direction in a given patch of sky, as we shall see in Sec.~{\ref{subsec:ang}}), they can never fully eliminate the presence of Hawking radiation, and the effective temperature will always change sign once they have entered into a region of the spacetime geometry where their local surface gravity [governed by radial gradient of the black hole's horizon function $\Delta$, Eq.~(\ref{eq:surface_gravity})] exceeds that of the Unruh emitter (or vice versa). This radiation in the radial direction can thus be regarded as ``real'' in the sense that it behaves in the same Lorentz-covariant way as any classical radiation detected by a free-faller would.

\subsection{\label{subsec:adi}Adiabaticity}

\begin{figure*}[t]
\centering
\begin{minipage}[l]{0.85\columnwidth}
  \includegraphics[width=\columnwidth]{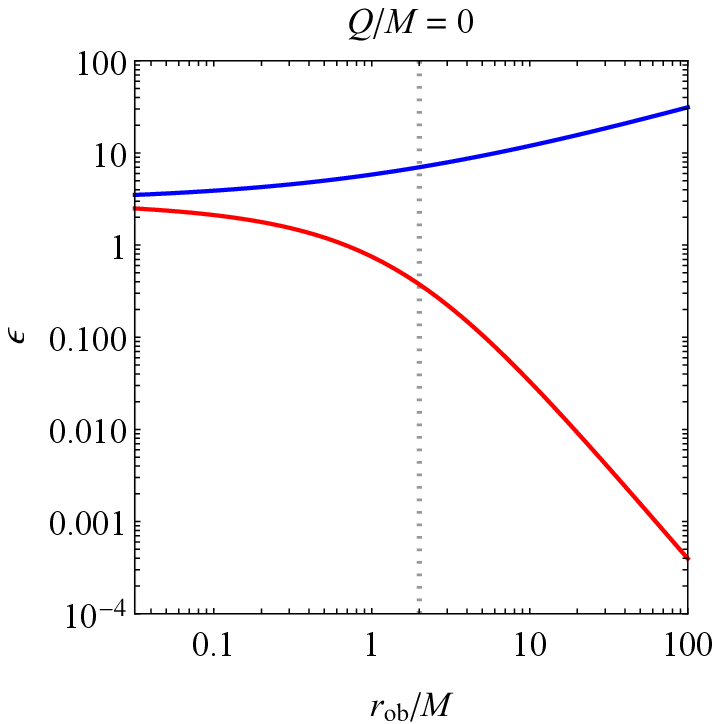}
\end{minipage}%
\hspace{1.5em}
\begin{minipage}[r]{0.85\columnwidth}
    \includegraphics[width=\columnwidth]{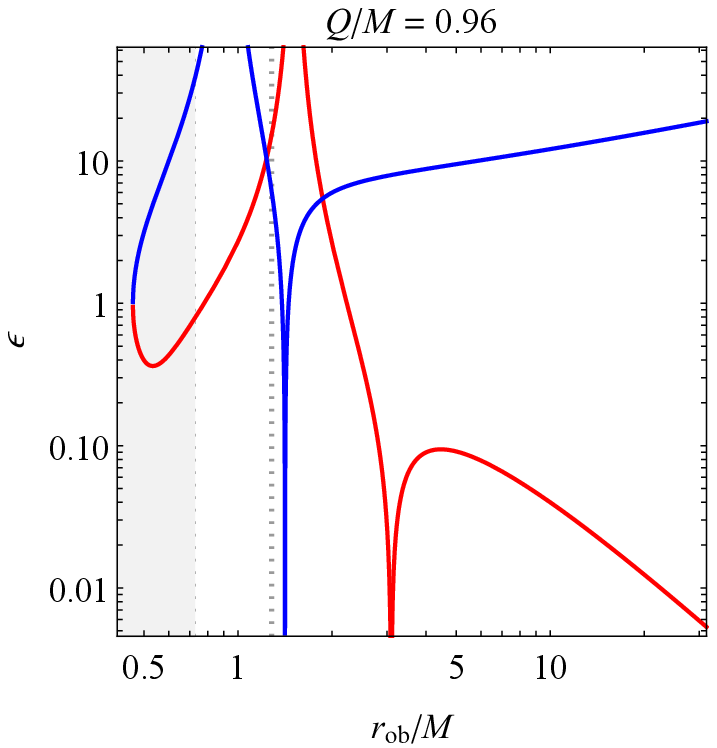}
\end{minipage}
\caption{Outgoing and ingoing adiabatic control functions $\epsilon^+$ (red curve) and $\epsilon^-$ (blue curve), respectively, as a function of an ingoing observer's position $r_\text{ob}$ for two choices of the Reissner-Nordstr\"om black hole charge $Q$, in units of the black hole mass $M$. As in Fig.~\ref{fig:kappa_RN}, the inner and outer horizons are shown with gray, dotted vertical lines, and the unphysical region below the inner horizon is grayed out.\label{fig:epsilon_RN}}
\end{figure*}

As mentioned in Sec.~\ref{subsec:foreff}, the identification of the effective temperature $\kappa$ with a thermal Hawking flux is strictly only valid in conjunction with the adiabatic condition, that $\kappa$ must remain approximately constant over enough e-folds of the arriving modes \cite{bar11a,bar11b}. This condition is quantified by the adiabatic control function $\epsilon$, which for radial modes in a static, spherically symmetric black hole can be written as
\begin{equation}\label{eq:epsilon_RN}
    \epsilon(r_\text{ob})\equiv\left|\frac{\dot{\kappa}}{\kappa^2}\right|=\left|\frac{\dot{r}_\text{ob}}{\kappa^2}\frac{d\kappa}{dr_\text{ob}}\right|.
\end{equation}
Whenever ${\epsilon\ll1}$, the adiabatic condition is satisfied and a thermal Hawking spectrum is perceived by the observer.

The exact analytic form of $\epsilon(r_\text{ob})$ for the Reissner-Nordstr\"om free-faller in the Unruh state is not too illuminating; nonetheless, several key features can be identified. As ${r_\text{ob}\to\infty}$, the adiabatic control function for the outgoing modes drops to zero (as anticipated to recover Hawking's original thermal calculation), except in the extremal case where $\kappa$ itself is already zero and $\epsilon$ therefore diverges. Similar diverging behavior in $\epsilon$ is observed whenever the effective temperature $\kappa$ vanishes, as a result of the $\kappa^2$ term in the denominator of Eq.~(\ref{eq:epsilon_RN}), since it is meaningless to define a thermal flux at zero temperature.

Based on the above observations, one might expect that $\epsilon$ would drop to zero whenever $\kappa$ diverges (e.g.\ when one observes outgoing modes at the inner horizon). However, the adiabatic control function at the inner horizon instead passes through a finite, nonzero value, which nonetheless is still usually smaller than unity for outgoing modes. Specifically, for an ingoing observer,
\begin{align}\label{eq:epsilon_RN_r-}
    &\lim_{r_\text{ob}\to r_-}\left\{\epsilon^+,\ \epsilon^-\right\}=\nonumber\\
    &\left\{\frac{r_+^2}{2(2M^2-Q^2)},\ \frac{5Q^2+4M\sqrt{M^2-Q^2}-3M^2}{M^2-Q^2}\right\}.
\end{align}
This equality technically only holds when ${Q\neq0}$; in the Schwarzschild case, instead of approaching unity, both $\epsilon^+$ and $\epsilon^-$ will asymptotically approach 3 (see the left panel of Fig.~\ref{fig:epsilon_RN}). But for ${Q>0}$, the value of $\epsilon^+$ at the inner horizon is always less than 1, and $\epsilon^-$ is always greater than 1. For large enough charge $Q$, Eq.~(\ref{eq:epsilon_RN_r-}) thus implies that the outgoing temperature should be approximately thermal for an ingoing observer close enough to the inner horizon. This behavior holds even (and especially) for ${Q/M=1}$, where the inflating negative temperature just above the merged horizons occurs in the black hole's exterior.

For reference, the behavior of ${\epsilon^+(r_\text{ob})}$ and ${\epsilon^-(r_\text{ob})}$ are plotted in Fig.~\ref{fig:epsilon_RN} for two of the same values of $Q$ used in Fig.~\ref{fig:kappa_RN}. One may observe that for many choices of $r_\text{ob}$, $\kappa$ behaves adiabatically and the thermal results fall into place. However, for much of the observer's trajectory, $\epsilon$ far exceeds unity, and deeper analysis is required, as examined in Sec.~\ref{sec:spe}.

One final technical point related to the discussion of adiabaticity is the comment made by the authors of Ref.~\cite{bar11a} that the effective temperature adiabaticity formalism described above is valid ``under mild technical assumptions.'' These assumptions are related to the more generalized, precise form of the adiabaticity condition, which assumes the existence of a finite quantity
\begin{equation}\label{eq:D}
    D\equiv\underset{n>0}{\text{sup}}\left[\frac{1}{(n+1)!}\frac{|\kappa^{(n)}|}{\kappa^{n+1}}\right]^{1/(n+1)}
\end{equation}
such that adiabaticity is implied by the condition ${2D^2\ll1}$ (instead of ${\epsilon\ll1}$). Usually, the ${n=1}$ term in the definition of Eq.~(\ref{eq:D}) dominates so that the quantity ${2D^2}$ is equivalent to the adiabatic control function $\epsilon$ of Eq.~(\ref{eq:epsilon_RN}). But in certain special cases, such as the dip observed in the blue curve in the right panel of Fig.~\ref{fig:epsilon_RN} as $\epsilon^-$ goes to zero just above the outer horizon, higher-$n$ terms in Eq.~(\ref{eq:D}) dominate. As a result, adiabaticity is not satisfied there, even though $\dot{\kappa}^-$ (and therefore $\epsilon^-$) vanishes.

\subsection{\label{subsec:ang}General viewing direction}

\begin{figure*}[t]
\centering
\begin{minipage}[l]{0.85\columnwidth}
  \includegraphics[width=\columnwidth]{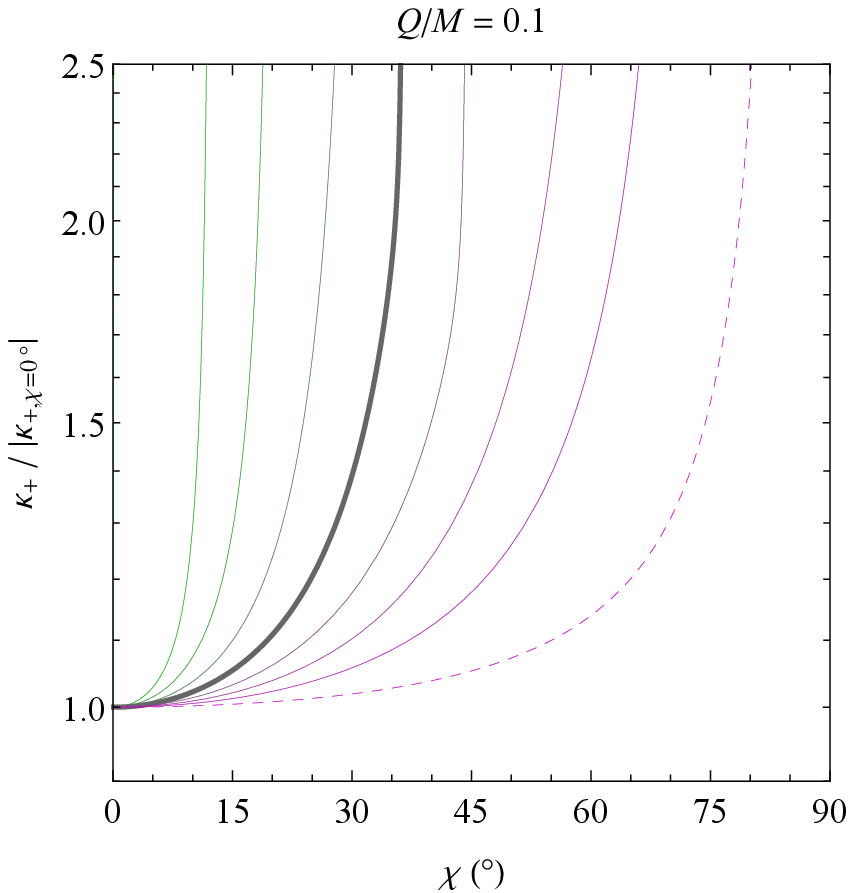}
\end{minipage}%
\hspace{1.5em}
\begin{minipage}[r]{0.85\columnwidth}
    \includegraphics[width=\columnwidth]{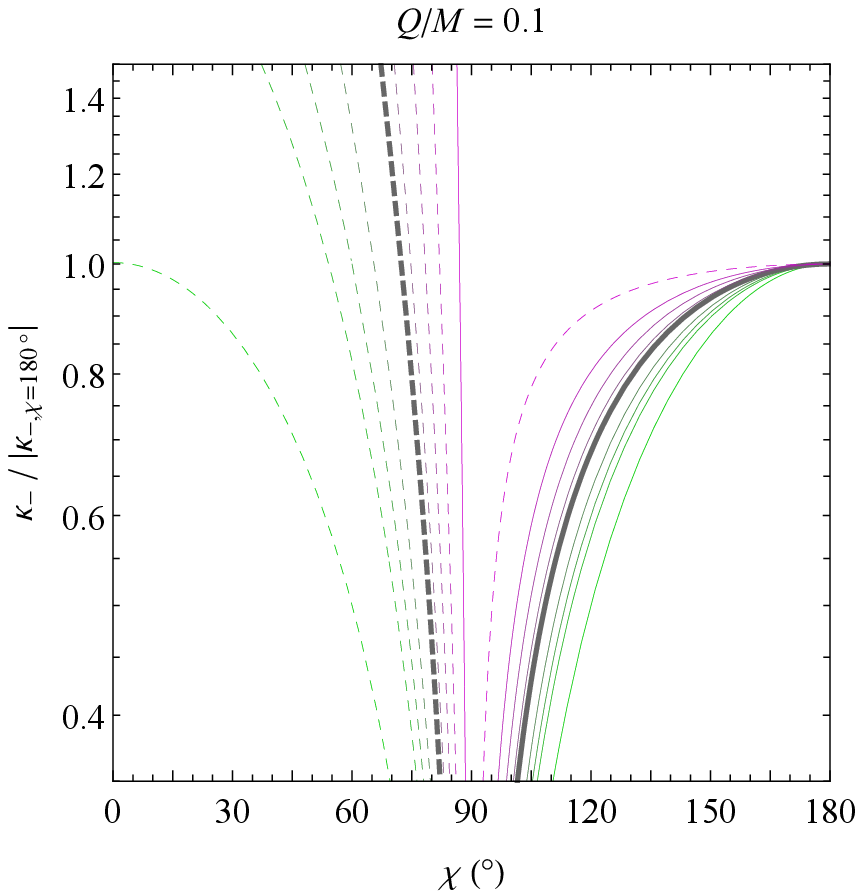}
\end{minipage}%
\vspace{1em}
\begin{minipage}[l]{0.85\columnwidth}
  \includegraphics[width=\columnwidth]{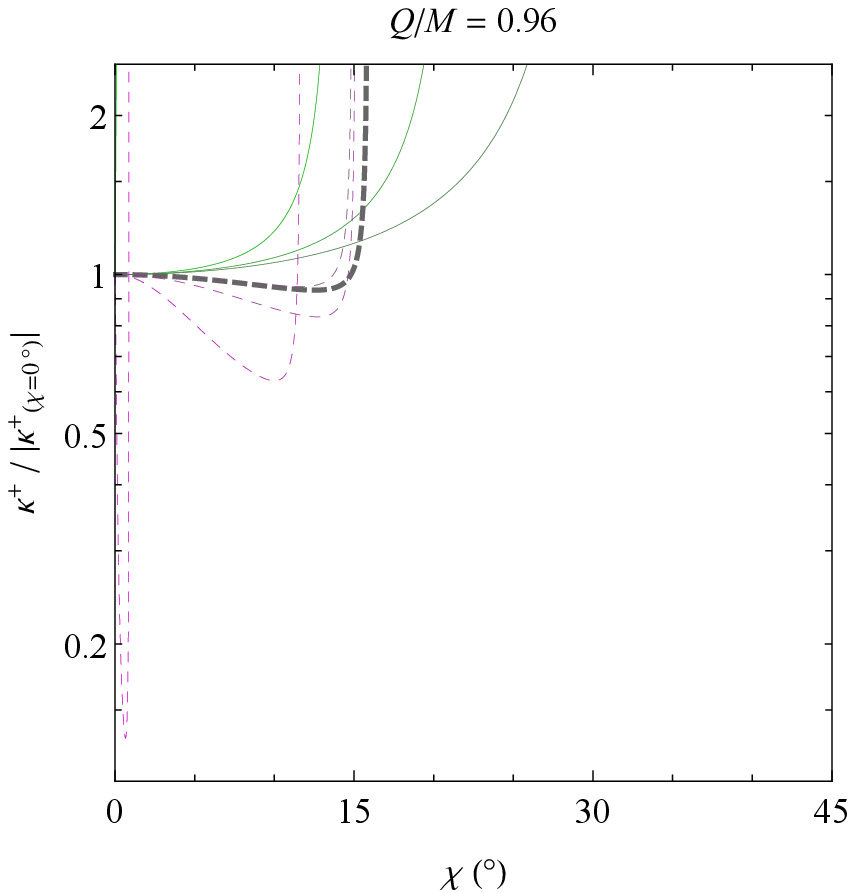}
\end{minipage}%
\hspace{1.5em}
\begin{minipage}[r]{0.85\columnwidth}
    \includegraphics[width=\columnwidth]{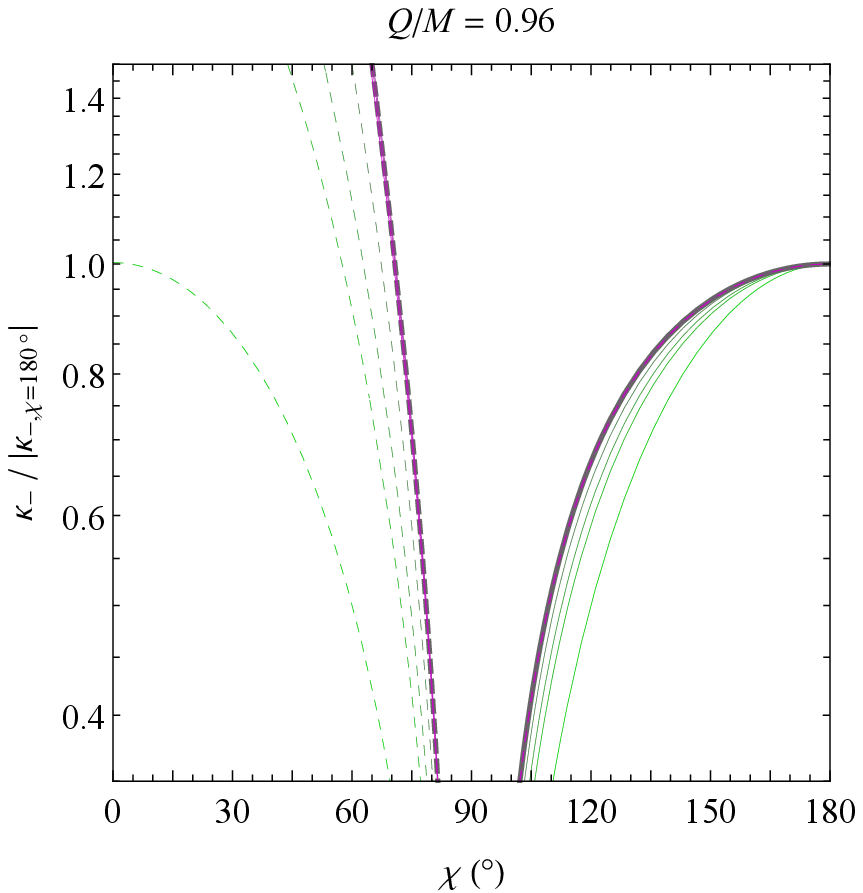}
\end{minipage}
\caption{Effective temperatures $\kappa^+$ (left two panels) and $\kappa^-$ (right two panels) seen by a radial, inertial, nonrotating observer falling from infinity to the left leg of the inner horizon, as a function of the observer's viewing angle $\chi$ on the sky. Curves from green to magenta indicate radiation observed at radii $r_\text{ob}\to\infty$, $8r_+$, $4r_+$, $2r_+$, $r_+$ (thick line), $r_-+0.5(r_+-r_-)$, $r_-+0.25(r_+-r_-)$, $r_-+10^{-1}(r_+-r_-)$, and $r_-+10^{-3}(r_+-r_-)$. All curves are normalized so that the magnitude of $\kappa^+$ or $\kappa^-$ for a given radius when looking, respectively, straight down (${\chi=0^\circ}$) or up (${\chi=180^\circ}$), is 1. Solid curves indicate positive values on the log plot, and dashed curves indicate negative values. \label{fig:kappa_chi}}
\end{figure*}

The results of Secs.~\ref{subsec:rad} and \ref{subsec:adi} apply to a radial infaller observing modes purely in the radial direction. Since the mass inflation instability involves radial focusing of all null geodesics, one may wonder whether the diverging acceleration seen by an infaller is confined to a single radial point on the sky.

The goal of this section is to provide a generalization of Eq.~(\ref{eq:kappa_chainrule}) to account for photons reaching the observer from any direction. The photon's 4-momentum will now include additional angular terms with the conserved quantity ${b\equiv k_\vartheta/k_t}$, the photon's impact parameter, which equals 0 for radial trajectories but in general can take any real value up to infinity. To translate $b$ into a viewing angle on the observer's sky, it suffices to define a single parameter $\chi$ that measures the angle in the observer's local tetrad frame between the radial direction and the direction the observer is facing. This viewing angle $\chi$ ranges from 0 degrees (facing radially inward toward the past horizon) to 180 degrees (facing radially outward toward the sky above, at past null infinity). For an observer with specific energy $E_\text{ob}$ at radius $r_\text{ob}$, the impact parameter $b$ is related to the viewing angle $\chi$ by \cite{ham18}
\begin{equation}
    b=\left|\frac{r_\text{ob}\sin\chi}{E_\text{ob}-\sqrt{E_\text{ob}^2-\Delta(r_\text{ob})}\cos\chi}\right|.
\end{equation}

The frequency $\omega$ measured in the frame of an observer ($\equiv\omega_\text{ob}$) or emitter ($\equiv\omega_\text{em}$) with specific energy $E$, normalized to the frequency $\omega_\infty$ seen at rest at infinity, then generalizes to
\begin{equation}\label{eq:omega_b}
    \frac{\omega}{\omega_\infty}=\frac{E\pm\sqrt{\left(E^2-\Delta\right)\left(1-b^2\Delta/r^2\right)}}{\Delta},
\end{equation}
where, as in the radial case, the upper (lower) sign applies to outgoing (ingoing) null rays. The calculation of $\kappa$ then follows as in the radial case, though great care must be taken to account for turnaround radii and ensure the correct sign for different viewing angles and observer positions.

Since the perception of particle production is highly dependent on the choice of observer, one must take care to make an appropriate choice depending on the context of the calculation. For example, an observer staring in a fixed direction $\chi$ as they fall inward is not the same as an observer staring at a single infalling emitter, whose position will constantly change in the observer's field of view. As argued in Ref.~\cite{ham18}, the choice of observer that will introduce the least amount of noninertial radiative effects (e.g.\ from the rotation of the observer's frame) and will reveal the most ``pure'' Hawking radiation is an observer staring in a fixed direction $\chi$. Such an observer will see a family of infalling emitters as they fall inward, with each emitter connected to the observer by a null path with the same phase.

If an observer stares at the sky above (corresponding to the ingoing Hawking sector, with a family of Unruh emitters at ${r_\text{em}\to\infty}$), the generalization of Eq.~(\ref{eq:kappa_chainrule}) to account for the frequency of Eq.~(\ref{eq:omega_b}) seen from any viewing angle $\chi$ is sufficient to satisfy the requirement from the previous paragraph of an inertial observer with fixed $\chi$. However, if the observer stares at the past horizon below them (corresponding to the outgoing Hawking sector, with a family of Unruh emitters at ${r_\text{em}\to r_+}$), the frequency seen by the emitter or the observer will diverge, as will the affine distance of the null geodesics connecting the two infallers. In order to ensure that the observer is seeing the same emitted in-modes as they follow along a geodesic staring in a fixed direction $\chi$, the emitted affine distance
\begin{equation}
    \lambda_\text{em}\equiv\omega_\text{em}\lambda=\omega_\text{em}\int_{r_\text{em}}^{r_\text{ob}}\frac{dr}{k^r}
\end{equation}
(where ${k^r\equiv dr/d\lambda}$ is the radial component of photon's coordinate-frame 4-momentum, given by Eq.~(80) of Ref.~\cite{ham18}) must be held constant. The resulting effective temperature then takes the form:
\begin{align}\label{eq:kappa_chainrule_b}
    \kappa&=-\frac{\partial}{\partial\tau_\text{ob}}\ln\left(\frac{\omega_\text{ob}\lambda}{\lambda_\text{em}}\right)\bigg|_{\chi,\lambda_\text{em}}\nonumber\\
    &=-\dot{r}_\text{ob}\left(\frac{\partial\ln\omega_\text{ob}}{\partial r_\text{ob}}\bigg|_{\chi}+\frac{\partial\ln\lambda}{\partial r_\text{ob}}\bigg|_{\chi}\right)-\dot{r}_\text{em}\frac{\omega_\text{ob}}{\omega_\text{em}}\frac{\partial\ln\lambda}{\partial r_\text{em}}\bigg|_{\chi},
\end{align}
where the derivatives of the affine distance (at constant $\chi$) can be expanded with the Leibniz integral rule:
\begin{subequations}
\begin{align}
    \frac{\partial\ln\lambda}{\partial r_\text{ob}}\bigg|_{\chi}&=\frac{1}{\lambda}\left(\frac{1}{k^r_\text{ob}}+\frac{\partial b}{\partial r_\text{ob}}\bigg|_{\chi}\int_{r_\text{em}}^{r_\text{ob}}\!\!\!dr\frac{\partial}{\partial b}\frac{1}{k^r}\right),\\
    \frac{\partial\ln\lambda}{\partial r_\text{em}}\bigg|_{\chi}&=-\frac{1}{\lambda k^r_\text{em}}.
\end{align}
\end{subequations}

The numerical solution to Eq.~(\ref{eq:kappa_chainrule_b}) for various values of $r_\text{ob}$ and $Q$ is shown in Fig.~\ref{fig:kappa_chi}. These plots show similar trends to that found in Ref.~\cite{ham18} for Schwarzschild black holes. First, the outgoing Hawking radiation seen from the past horizon (left two panels) is actually weakest in the radial direction (except when the observer is very close to the inner horizon). As $\chi$ increases from 0$^\circ$ and the observer looks farther away from the center of the black hole's shadow marking where the past horizon would be, $\kappa^+$ increases until it diverges at the edge of the shadow.\footnote{This divergence is an artifact of the unphysical metric used; for an astrophysical black hole formed by gravitational collapse a finite time in the past, the Hawking radiation would still exponentially limb brighten but would remain finite before dropping to zero outside of the black hole's shadow \cite{ham18}.} As the observer falls closer and closer to the inner horizon, the area of sky across which Hawking radiation is visible becomes larger (in conjunction with the growing apparent size of the black hole's shadow), and the Hawking radiation becomes more and more isotropic across the surface of the shadow. But once the observer falls close enough to the inner horizon, the apparent black hole size begins to decrease as the Hawking area shrinks to a small patch of sky ahead of the observer (this effect is most apparent in the lower left panel of Fig.~\ref{fig:kappa_chi}, but even in the upper left panel, additional curves for smaller radii $r_\text{ob}$ would begin to shrink since the maximum angle $\chi$ shifts down to 0$^\circ$ as ${r\to r_-}$).

When the black hole's charge $Q$ is nonzero, the main effect on the outgoing effective temperature at arbitrary viewing angle is the same result found in Sec.~\ref{subsec:rad}; namely, an observer close enough to the inner horizon will see a negative $\kappa^+$, corresponding to modes that are exponentially blueshifting instead of redshifting. The higher the charge $Q$, the farther out in physical space this blueshifting zone becomes, until it extends beyond the outer horizon and reaches infinity in the extremal case (as already seen in the radial case of Fig.~\ref{fig:kappa_RN_negregion}).

Similarly, ingoing Hawking radiation seen from an observer looking up at the sky above (right two panels of Fig.~\ref{fig:kappa_chi}) reproduces the same behavior found in Ref.~\cite{ham18} for the Schwarzschild case, with minimal modifications when $Q$ is nonzero. The rate of redshifting in the upper hemisphere is strongest when the observer looks straight up to the sky (in the outward radial direction, ${\chi=180^\circ}$), and $\kappa^-$ changes sign at $90^\circ$, reflecting the fact that the infaller is accelerating away from the sky above (so that the upper hemisphere is redshifting) and accelerating toward the black hole below (so that the lower hemisphere is blueshifting).

However, as with the outgoing effective temperature, the ingoing effective temperature changes sign once the observer falls close enough to the inner horizon [seen, e.g., with the dashed pink line at $r_\text{ob}={r_-+10^{-3}(r_+-r_-)}$ on the right half of the top right panel of Fig.~\ref{fig:kappa_chi}], so that the upper hemisphere is blueshifting and the lower hemisphere is redshifting. But unlike the outgoing radiation, the sign change in the ingoing effective temperature is restricted only to infallers within the event horizon, regardless of the value of $Q$. 

Aside from the sign reversal in every direction for observers close enough to the inner horizon, the main contribution that an addition of charge has on the angular distribution of Hawking radiation (for both $\kappa^+$ and $\kappa^-$) is to smooth out the perceived temperature gradients across the sky\textemdash the higher the charge $Q$, the less sharp the temperature cutoff is at the black hole shadow's boundaries, and therefore the less isotropic the temperature is across the observer's field of view for a given distance above the inner horizon.

\subsubsection{Dependence on the observer's energy}

\begin{figure*}[t]
\centering
\begin{minipage}[l]{0.95\columnwidth}
  \includegraphics[width=\columnwidth]{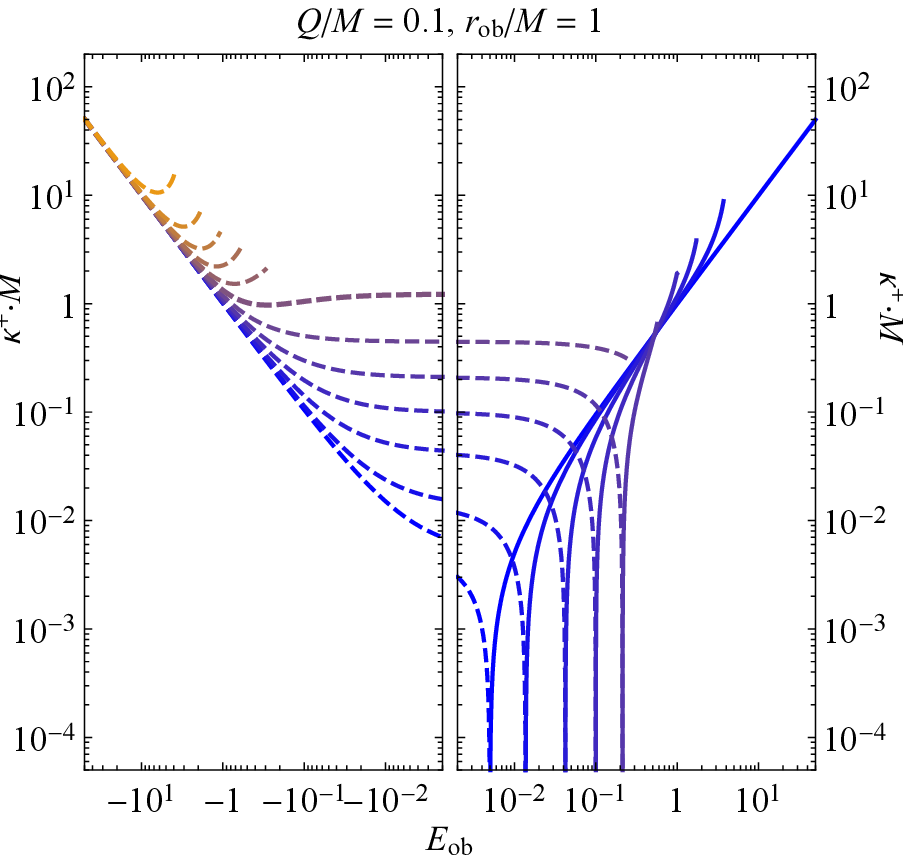}
\end{minipage}%
\hspace{1.5em}
\begin{minipage}[r]{0.95\columnwidth}
    \includegraphics[width=\columnwidth]{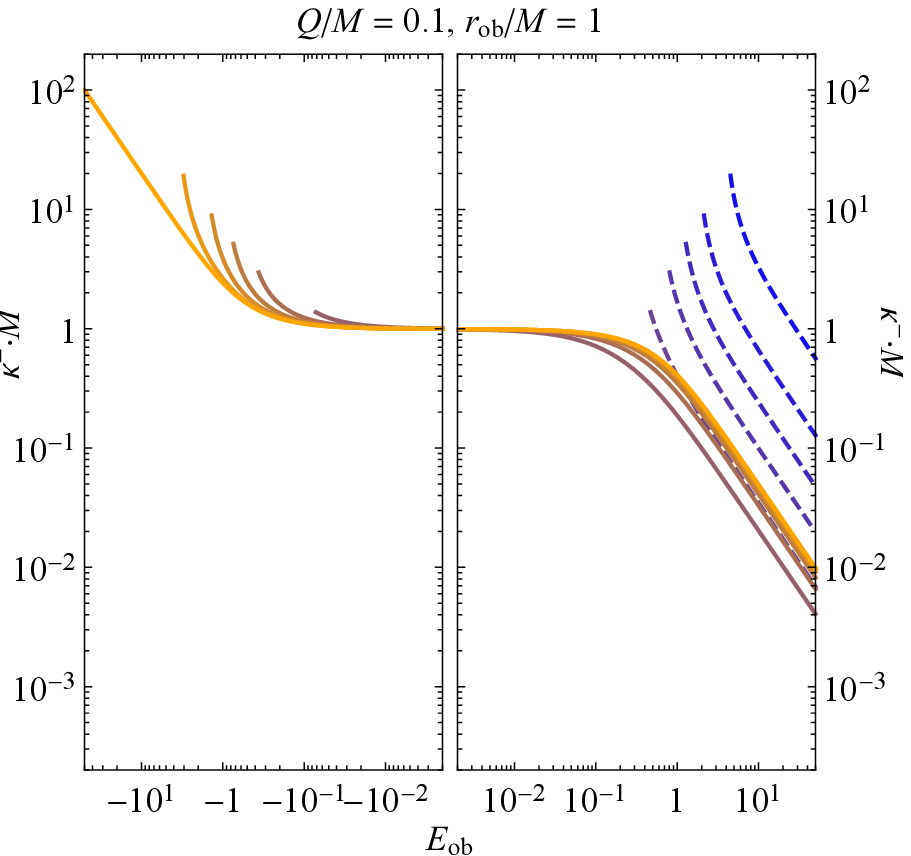}
\end{minipage}%
\vspace{1em}
\begin{minipage}[l]{0.95\columnwidth}
  \includegraphics[width=\columnwidth]{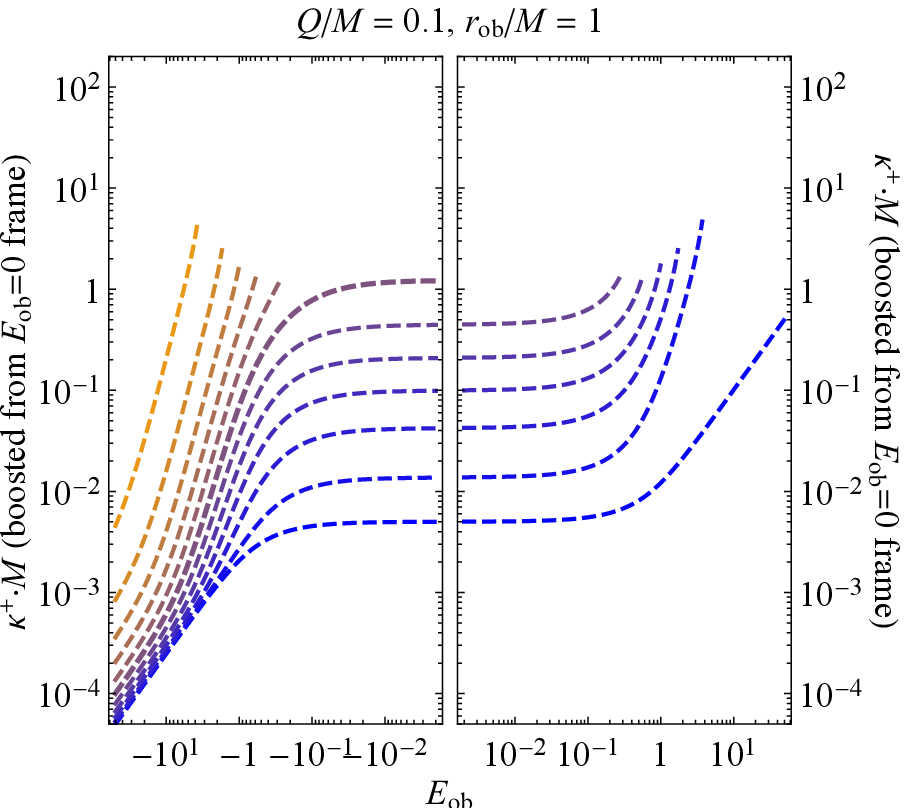}
\end{minipage}%
\hspace{1.5em}
\begin{minipage}[r]{0.95\columnwidth}
    \includegraphics[width=\columnwidth]{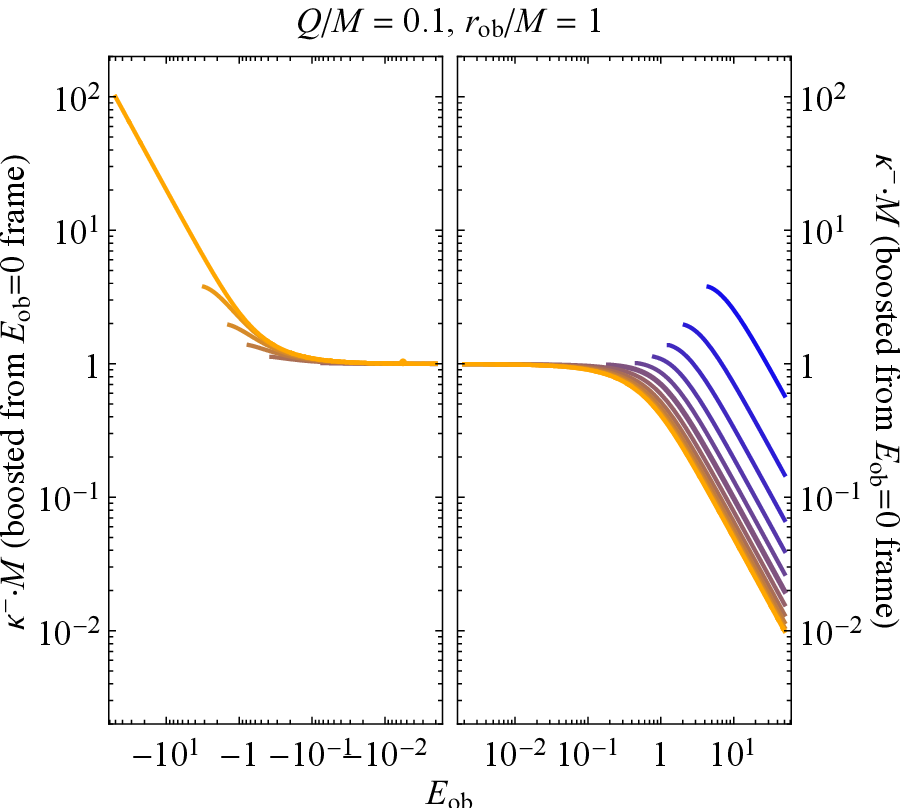}
\end{minipage}
\caption{Effective temperatures $\kappa^+$ (left plots) and $\kappa^-$ (right plots) in units of $M^{-1}$ as a function of the observer's specific energy $E_\text{ob}$, for various choices of the observer's viewing direction $\chi$, with intervals of 15$^\circ$ from $\chi=0^\circ$ (blue) to $\chi=180^\circ$ (orange) (note that the left plots contain no $\chi=180^\circ$ curves and the right plots contain no $\chi=0^\circ$ curves). Solid curves indicate positive values and dashed curves indicate negative values. The black hole's charge-to-mass ratio is ${Q/M=0.1}$, and the radiation is seen from an observer halfway between the inner and outer horizons, at ${r_\text{ob}/M=1}$. The upper two plots show the effective temperatures calculated from Eq.~(\ref{eq:kappa_chainrule_b}) directly as a function of $E_\text{ob}$, while the lower two plots calculate the effective temperatures only for ${E_\text{ob}=0}$ and infer the effective temperatures at other observer energies by Lorentz-boosting to the appropriate frame.\label{fig:kappa_chi_E}}
\end{figure*}

The dependence of the ingoing and outgoing effective temperatures $\kappa^-$ and $\kappa^+$ on the observer's specific energy $E_\text{ob}$ is shown in the upper two plots of Fig.~\ref{fig:kappa_chi_E}. These plots only show one choice of black hole charge (${Q/M=0.1}$) and observer position ($r_\text{ob}/M=1$) so that the relevant qualitative trends can be observed.

As a check on the consistency of the upper two plots in Fig.~\ref{fig:kappa_chi_E}, one can find that the presence or absence of different constant-$\chi$ curves at different observer energies exactly matches the position of the black hole silhouette in the observer's field of view. For example, for a black hole with ${Q/M=0.1}$, an observer at ${r_\text{ob}/M=1}$ with ${E_\text{ob}=1}$ will see the past horizon below them (the black hole's ``shadow'') spanning from $\chi=0^\circ$ to its border at approximately $\chi\approx53.2^\circ$, and in both upper plots at ${E_\text{ob}=1}$, the radiation $\kappa^-$ from the sky exists only for ${\chi>53.2^\circ}$ while the radiation $\kappa^+$ from the horizon exists only for ${\chi<53.2^\circ}$. This holds true for all observer energies\textemdash as an observer is Lorentz-boosted to ${E_\text{ob}\to\infty}$, the past horizon shrinks to a single point below them, and as they are boosted in the other direction (${E_\text{ob}\to-\infty}$), the sky shrinks to a single point above them.

The lower two plots of Fig.~\ref{fig:kappa_chi_E} give a further check on the consistency of the formalism and help to show the degree to which the effective temperatures satisfy Lorentz covariance. As the observer's energy $E_\text{ob}$ changes, the observer is effectively Lorentz-boosting to a different frame, even though no restriction was imposed \emph{a priori} for the effective temperature to transform under the Lorentz group. As a test, the lower two plots of Fig.~\ref{fig:kappa_chi_E} start with the same calculations of $\kappa^+$ and $\kappa^-$ at ${E_\text{ob}=0}$, but instead of varying $E_\text{ob}$ in Eq.~(\ref{eq:kappa_chainrule_b}) to find the effective temperature at other observer energies, a Lorentz boost is applied to the observer and matched to the different energies. When beginning in the ${E_\text{ob}=0}$ frame, an interior observer boosted to a frame where they have energy $E_\text{ob}'$ will possess the Lorentz factor
\begin{equation}
    \gamma=\sqrt{\frac{E_\text{ob}'^2-\Delta}{\Delta}}.
\end{equation}
Such a boost will entail two important effects. First, the effective temperature will be Doppler-shifted by the frequency factor $\omega_\text{ob}$ from Eq.~(\ref{eq:omega_b}), normalized to the frequency seen in the ${E_\text{ob}=0}$ frame. And second, the observer's field of view will experience relativistic aberration, such that photons arriving at an angle $\chi$ for the ${E_\text{ob}=0}$ observer will be shifted to the angle
\begin{equation}
    \chi'=\cos^{-1}\left(\frac{\cos\chi+\beta}{1+\beta\cos\chi}\right)
\end{equation}
in the boosted frame (where ${\beta=\sqrt{1-\gamma^{-2}}}$ is the observer's speed). If the Hawking radiation seen by the observer behaved purely classically and in a Lorentz-covariant fashion, the upper two plots of Fig.~\ref{fig:kappa_chi_E} would exactly match their lower counterparts.

As anticipated by the radial case (see Sec.~\ref{subsubsec:kappa_E}), the Hawking radiation seen from the sky above ($\kappa^-$, upper right plot in Fig.~\ref{fig:kappa_chi_E}) in every direction varies reciprocally with $E_\text{ob}$ as the observer's energy asymptotically increases and varies approximately linearly as ${E_\text{ob}\to-\infty}$. Such a behavior is similar to what is expected for a Lorentz-boosted observer as in the lower right plot of Fig.~\ref{fig:kappa_chi_E}. And just as in the radial case, changing the observer's specific energy $E_\text{ob}$ for fixed $\chi$ will never change the sign of $\kappa^-$. The ingoing effective temperature is always zero when ${\chi=90^\circ}$, always positive (with this specific choice of observer halfway between the outer and inner horizons) for larger $\chi$, and always negative for smaller $\chi$. Such a delineation can be noticed in the upper right plot of Fig.~\ref{fig:kappa_chi_E} from the fact that the ${\chi<90^\circ}$ curves (blue) are always negative (dashed), while the ${\chi>90^\circ}$ curves (orange) are always positive (solid). This behavior is a consequence of forcing the observer to stare in a fixed direction; such an infaller will classically always see null geodesics from infinity blueshifting below them (when ${\chi<90^\circ}$) and redshifting above them as they decrease their radius.

What about the outgoing Hawking radiation from the horizon? As shown in the upper left plot of Fig.~\ref{fig:kappa_chi_E}, an interior observer can change the sign of $\kappa^+$ by changing their energy $E_\text{ob}$ enough. When ${E_\text{ob}=1}$, the results of the upper left panel of Fig.~\ref{fig:kappa_chi} are reproduced; namely, a positive-temperature horizon is seen with brighter radiation at the edges (i.e.\ larger $\kappa^+$ for larger $\chi$). However, as the observer boosts to smaller and smaller energies, the temperature at the ever-growing edge of the horizon begins to decrease until it drops below zero. The negative-temperature outer portion of the black hole's shadow then begins to grow inward until the entire horizon has a negative temperature, once again with the largest magnitude at the edges. Though only one specific case is shown, an outgoing (i.e.\ negative-energy) observer in a black hole's interior will always see a completely negative-temperature horizon below them.

One way that the upper left plot of Fig.~\ref{fig:kappa_chi_E} differs from the results of Sec.~\ref{subsubsec:kappa_E} (and from the lower left plot of Fig.~\ref{fig:kappa_chi_E}) is that $\kappa^+$ diverges linearly as ${E_\text{ob}\to-\infty}$ instead of dropping to zero. As a reminder, the difference in the calculation done here versus that of previous sections is that here the affine distance is kept constant so that the family of emitters seen by the observer will always have the same phase, since the emitted wave's frequency appears to diverge as the emitter is taken to the horizon. Evidently such a restriction has a big impact not just in the evaluation of the horizon temperature for nonzero $\chi$, but also for the evaluation of the horizon temperature for negative observer energies, even when ${\chi=0^\circ}$.

Finally, let us briefly give special attention to the case of an interior observer with ${E_\text{ob}=0}$. Classically, such an observer will begin at the event horizon seeing nothing but the past horizon in all directions, excepting a vanishingly small patch of sky directly above them at ${\chi=180^\circ}$. Then, as they fall inwards, the sky above them will grow until it almost takes up a full hemisphere of the observer's field of view, after which the sky will quickly collapse back to a single point as the horizon grows. Semiclassically, in Sec.~\ref{subsubsec:kappa_rm} it was argued that the Hawking radiation in the ${E_\text{ob}=0}$ frame diverges as ${z_\text{ob}^{-1/2}}$ as an observer approaches the inner horizon looking both up (${\chi=180^\circ}$) and down (${\chi=0^\circ}$). What happens in other directions?

When ${E_\text{ob}=0}$, the effective temperature from the sky above becomes isotropic and simplifies considerably:
\begin{equation}\label{eq:kappa-_chi_r-}
    \lim_{E_\text{ob}\to0}\kappa^-(\chi)=\frac{1}{2\sqrt{-\Delta}}\frac{d\Delta}{dr_\text{ob}}.
\end{equation}
This radiation extends across the entire sky visible to the observer, from $\chi=180^{\circ}$ to the edge of the black hole shadow at
\begin{equation}
    \cos\chi=-\left[1-\frac{\Delta(r)}{r^2}\frac{r_c^2}{\Delta(r_c)}\right]^{-1/2},
\end{equation}
where ${r_c\equiv\frac{3M}{2}\left(1+\sqrt{1-\frac{8Q^2}{9M^2}}\right)}$ is the critical radius of the photon sphere. This isotropicity can be seen by the convergence of all the curves in the right plots of Fig.~\ref{fig:kappa_chi_E} as ${E_\text{ob}\to0}$.

The effective temperature $\kappa^+$ from the horizon does not take on a simple analytic form like $\kappa^-$ does, but its dependence on $\chi$ for an observer with ${r_\text{ob}/M=1}$ can be ascertained from the left plots of Fig.~\ref{fig:kappa_chi_E}. For various charges $Q$ and observer positions $r_\text{ob}$, the effective temperature is usually negative in all directions, with the smallest magnitude for $\kappa^+$ occurring when looking straight downward (${\chi=0^\circ}$). Notably, as the observer reaches the inner horizon, while the temperature $\kappa^-$ from Eq.~(\ref{eq:kappa-_chi_r-}) diverges isotropically as ${(-\Delta)^{-1/2}}$ (and therefore as $z_\text{ob}^{-1/2}$), the temperature $\kappa^+$ from the horizon also diverges as ${(-\Delta)^{-1/2}}$, with an even stronger divergence when ${\chi>0^\circ}$.

\section{\label{sec:spe}Bogoliubov spectrum}

Since a variety of choices for the observer position $r_\text{ob}$ and black hole charge $Q$ lead to a nonadiabatic effective temperature function, one may wonder how much trust can be placed on the physical validity of the results of Sec.~\ref{sec:redshift}. As has been argued, even if the Hawking spectrum is nonthermal, there should in general still be particle production whenever $\kappa$ is nonzero. To verify this claim, here we will perform a full wave mode analysis to find the particle spectrum seen by an infaller in the locations where the Klein-Gordon equation simplifies enough for such a calculation to be performed.

\subsection{Derivation}

Consider the Bogoliubov coefficients between the vacuum state of an Unruh emitter and that of a freely falling observer in a Reissner-Nordstr\"om spacetime. In any spacetime with metric $g_{\mu\nu}$, a canonically quantized, massless scalar field $\Phi$ will satisfy the Klein-Gordon wave equation
\begin{equation}\label{eq:waveeq_phi}
    \frac{1}{\sqrt{-g}}\frac{\partial}{\partial x^\mu}\left(\sqrt{-g}\ g^{\mu\nu}\frac{\partial\Phi}{\partial x^\nu}\right)=0.
\end{equation}
Motivated by the spacetime's symmetries, we choose to decompose this field $\Phi$ into a complete set of modes $\phi_{\omega\ell m}$, each accompanied by creation and annihilation operators $a^\dagger$ and $a$:
\begin{equation}\label{eq:modedecomp}
    \Phi=\int_0^\infty d\omega\sum_{\ell=0}^\infty\sum_{m=-\ell}^\ell\left(\phi_{\omega\ell m}a_{\omega\ell m}+\phi^*_{\omega\ell m}a^\dagger_{\omega\ell m}\right).
\end{equation}
If these modes are separated as
\begin{equation}\label{eq:RN_modesep}
    \phi_{\omega\ell m}=\frac{f_{\omega\ell}(t,r)Y_{\ell m}(\theta,\varphi)}{r\sqrt{4\pi\omega}},
\end{equation}
then Eq.~(\ref{eq:waveeq_phi}) implies that $Y_{\ell m}$ will satisfy the spherical harmonic equation, while $f_{\omega\ell}$ must satisfy
\begin{equation}\label{eq:waveeq_RN_tr}
    \frac{\partial^2f_{\omega\ell}}{\partial r^{*2}}-\frac{\partial^2f_{\omega\ell}}{\partial t^2}=\Delta\left[\frac{\ell(\ell+1)}{r^2}+\frac{1}{r}\frac{d\Delta}{dr}\right]f_{\omega\ell}.
\end{equation}

The annihilation operators ${a_{\omega\ell m}}$ of Eq.~(\ref{eq:modedecomp}) define the vacuum state of the observer:
\begin{equation}
    a_\text{ob}|0_\text{ob}\rangle=0
\end{equation}
(for convenience, the mode indices $\omega$, $\ell$, and $m$ will hereafter be suppressed as needed). However, $\Phi$ could just as easily be decomposed into any other complete set of modes $\bar{\omega}$, $\bar{\ell}$, and $\bar{m}$, so a similar decomposition can be used to define an emitter's vacuum state as
\begin{equation}
    a_\text{em}|0_\text{em}\rangle=0.
\end{equation}
The two vacuum states are related by a Bogoliubov transformation through the coefficients $\alpha^{\omega\ell m}_{\bar{\omega}\bar{\ell}\bar{m}}$ and $\beta^{\omega\ell m}_{\bar{\omega}\bar{\ell}\bar{m}}$ (and note that there should properly be a sum of two integrals for the emitter's ingoing and outgoing states, which are omitted here for simplicity):
\begin{equation}
    a_\text{ob}=\int_0^\infty d\bar{\omega}\sum_{\bar{\ell}=0}^\infty\sum_{\bar{m}=-\bar{\ell}}^{\bar{\ell}}\left(\alpha\ a_\text{em}+\beta^*a^{\dagger}_\text{em}\right).
\end{equation}
It is then straightforward to show \cite{bir82} that the vacuum expectation value of the observer's number operator in the emitter's vacuum state is related to the Bogoliubov coefficient $\beta$:
\begin{align}\label{eq:bog}
    \langle0_\text{em}|a^{\dagger}_\text{ob}&a_\text{ob}|0_\text{em}\rangle=\int_0^\infty d\bar{\omega}\sum_{\bar{\ell}=0}^\infty\sum_{\bar{m}=-\bar{\ell}}^{\bar{\ell}}\left|\beta\right|^2\nonumber\\
    &=\int_0^\infty d\bar{\omega}\sum_{\bar{\ell}=0}^\infty\sum_{\bar{m}=-\bar{\ell}}^{\bar{\ell}}\left|\langle\phi_\text{em}|\phi^*_\text{ob}\rangle\right|^2,
\end{align}
where bra-ket notation denotes the Lorentz-invariant Klein-Gordon inner product, which consists of a 3D integral over an arbitrary spacelike Cauchy hypersurface $\Sigma$ that terminates at spacelike infinity and is orthogonal to a future-directed unit vector $n^\mu$:
\begin{equation}\label{eq:inner_product}
    \langle\phi_1|\phi_2\rangle\equiv-i\int_\Sigma d\Sigma\ n^\mu\sqrt{-g_\Sigma}\ \phi_1\overset{\leftrightarrow}{\partial}_\mu\phi_2^*.
\end{equation}
To determine the expected particle number seen by an observer, one thus needs only to specify the observer's and emitter's modes (usually via a set of boundary conditions), propagated through the spacetime via the wave equation so that they coincide on some Cauchy hypersurface.

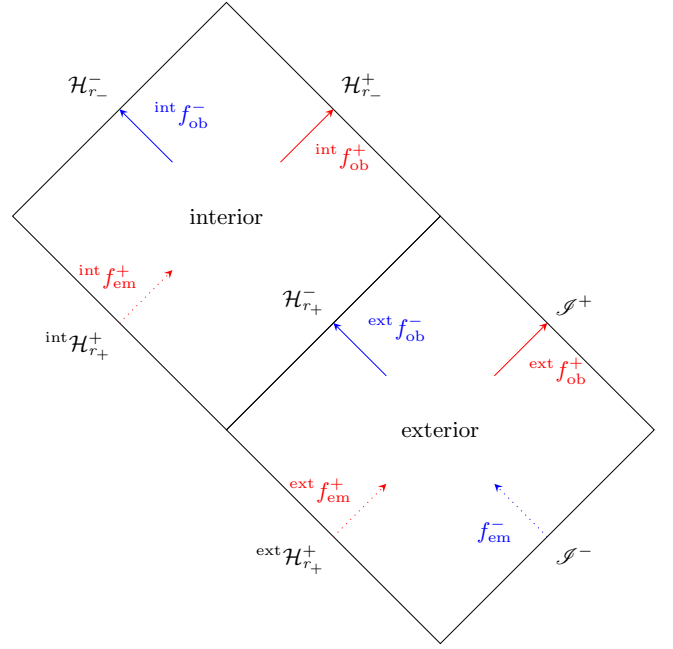
\begin{figure}[t]
\resizebox{\columnwidth}{!}{%
\begin{tikzpicture}
\def\x{1.5cm} % 1/4th the distance from top to bottom of a diamond
\node (ext)   at (\x,-\x) {exterior};
\node (int)   at (-\x,\x) {interior};
\path  % Four corners of top-left diamond
  (int) +(90:2*\x)  coordinate  (inttop)
        +(-90:2*\x) coordinate  (intbot)
        +(0:2*\x)   coordinate  (intright)
        +(180:2*\x) coordinate  (intleft);
\draw (intleft) -- 
          node[midway, above left] {$\cal{H}_{\textit{r}_-}^-$}
      (inttop) --
          node[midway, above right] {$\cal{H}_{\textit{r}_-}^+$}
      (intright) -- (intbot) --
          node[midway, below left] {${}^{\text{int}}\cal{H}_{\textit{r}_+}^+$}
      (intleft) -- cycle;
\path % Four corners of the bottom-right diamond
   (ext) +(90:2*\x)  coordinate%[label=75:$i^+$]
            (exttop)
         +(-90:2*\x) coordinate%[label=-90:$i^-$]
            (extbot)
         +(180:2*\x) coordinate
            (extleft)
         +(0:2*\x)   coordinate%[label=0:$i^0$]
            (extright);
\draw  (extleft) -- 
          node[midway, above left] {$\cal{H}_{\textit{r}_+}^-$}
    (exttop) --
          node[midway, above right] {$\mathscr{I}^+$}
    (extright) -- 
          node[midway, below right] {$\mathscr{I}^-$}
    (extbot) --
          node[midway, below left] {${}^{\text{ext}}\cal{H}_{\textit{r}_+}^+$}
    (extleft) -- cycle;
% Emitter mode arrows
\draw [dotted,red,-stealth](0,-2*\x) -- 
          node[midway, above left] {${}^\text{ext}f^+_\text{em}$}
    +(45:0.7*\x);
\draw [dotted,red,-stealth](-2*\x,0) --
          node[midway, above left] {${}^\text{int}f^+_\text{em}$}
    +(45:0.7*\x);
\draw [dotted,blue,-stealth](2*\x,-2*\x) --
          node[midway, below left] {$f^-_\text{em}$}
    +(135:0.7*\x);
% Observer mode arrows
\draw [red,stealth-](2*\x,0) -- 
          node[midway, below right] {${}^\text{ext}f^+_\text{ob}$}
    +(-135:0.7*\x);
\draw [red,stealth-](0,2*\x) --
          node[midway, below right] {${}^\text{int}f^+_\text{ob}$}
    +(-135:0.7*\x);
\draw [blue,stealth-](-2*\x,2*\x) --
          node[midway, above right] {${}^\text{int}f^-_\text{ob}$}
    +(-45:0.7*\x);
\draw [blue,stealth-](0,0) --
          node[midway, above right] {${}^\text{ext}f^-_\text{ob}$}
    +(-45:0.7*\x);
\end{tikzpicture}
}%
\caption{Penrose diagram showing the various boundaries for a Reissner-Nordstr\"om black hole on which modes are defined with nonzero values. Past (future) null infinity is labeled $\mathscr{I}^-$ ($\mathscr{I}^+$), the outer (inner) horizons are labeled $\cal{H}_{\textit{r}_+}$ ($\cal{H}_{\textit{r}_-}$), and the superscripts $+$ ($-$) everywhere indicate whether modes traveling across a surface are outgoing (ingoing). The boundary conditions for the emitter's (observer's) modes at the locations of the dotted (solid) lines can then be propagated (backpropagated) numerically using the wave equation to define the modes throughout the entire spacetime.\label{fig:RN_penrose}}
\end{figure}

The Unruh emitter's ingoing ($-$) and outgoing ($+$) modes are defined with the following boundary conditions at past null infinity $\mathscr{I}^-$ and the past horizon $\cal{H}^+_{\textit{r}_+}\equiv{}^{\text{int}}\cal{H}^+_{\textit{r}_+}\cup{}^{\text{ext}}\cal{H}^+_{\textit{r}_+}$ [here $f$ is defined as in Eq.~(\ref{eq:RN_modesep}), with $\omega\ell$ indices dropped for convenience]:
\begin{equation}\label{eq:RN_f_em+}
    f^+_\text{em}\to
    \begin{cases}
        0,&\mathscr{I}^-\\
        \text{e}^{-i\omega U},&\cal{H}^+_{\textit{r}_+}
    \end{cases},
\end{equation}
\begin{equation}\label{eq:RN_f_em-}
    f^-_\text{em}\to
    \begin{cases}
        \text{e}^{-i\omega(t+r^*)},&\mathscr{I}^-\\
        0,&\cal{H}^+_{\textit{r}_+}
    \end{cases},
\end{equation}
where $U$ is the outgoing Kruskal-Szekeres coordinate, defined in terms of the event horizon's surface gravity $\varkappa_+$ from Eq.~(\ref{eq:surface_gravity}) by
\begin{equation}
    U\equiv
    \begin{cases}
        -\text{e}^{-\varkappa_+(t-r^*)}/\varkappa_+,&r_+\leq r<\infty\\
        +\text{e}^{-\varkappa_+(t-r^*)}/\varkappa_+,&r_-\leq r<r_+
    \end{cases}.
\end{equation}
The relevant surfaces to which these boundary conditions correspond are shown schematically with dotted arrows in the Penrose diagram of Fig.~\ref{fig:RN_penrose}. Note that, as shown in the diagram, the outgoing modes can be further split into a pair of substates via $f^+\equiv{\left({}^\text{int}f^+\right)\cup\left({}^\text{ext}f^+\right)}$, each of whose boundary conditions are zero except on their respective null surfaces. As argued in Sec.~\ref{subsec:vac}, the modes of Eqs.~(\ref{eq:RN_f_em+}) and (\ref{eq:RN_f_em-}) are precisely those which are positive frequency with respect to the proper time of a freely falling observer skimming asymptotically close to those surfaces. The modes $f^\pm_\text{em}$ can then be extended to the entire spacetime by solving the wave Eq.~(\ref{eq:waveeq_RN_tr}).

Similarly, the observer's ingoing ($-$) and outgoing ($+$) modes can be defined via boundary conditions, in this case on the future null hypersurfaces. At future null infinity, the outgoing modes are positive frequency with respect to the outgoing Eddington-Finkelstein coordinate ${u\equiv t-r^*}$, since an observer asymptotically close to that surface will define positive frequency with respect to that coordinate (as argued in Sec.~\ref{subsec:vac}). The natural question is then how this vacuum state should be extended to the interior of the black hole. In studies of analogous acoustic black hole systems \cite{and13,bal19}, these interior modes are also defined with respect to the Eddington-Finkelstein coordinates, in part because the inner horizon of those systems is mimicked by a physically infinite asymptotic regime. For the Reissner-Nordstr\"om spacetime, an infaller will not reach an asymptotically steady state at the inner horizon; however, they will approach an asymptotic regime (albeit a transient one) where ${\Delta\to0}$ and the scattering potential of Eq.~(\ref{eq:waveeq_RN_tr}) vanishes. In the regime where this potential vanishes, as shown in Sec.~\ref{subsec:vac}, freely falling observers experience a proper time proportional to the Eddington-Finkelstein coordinates:
\begin{equation}\label{eq:RN_f_ob+}
    f^+_\text{ob}\to
    \begin{cases}
        \text{e}^{-i\omega(t-r^*)},&\mathscr{I}^+\\
        \text{e}^{-i\omega(r^*-t)},&\cal{H}^+_{\textit{r}_-}\\
        0,&\cal{H}^-_{\textit{r}_-}
    \end{cases},
\end{equation}
\begin{equation}\label{eq:RN_f_ob-}
    f^-_\text{ob}\to
    \begin{cases}
        0,&\mathscr{I}^+\cup\cal{H}^+_{\textit{r}_-}\\
        \text{e}^{-i\omega(r^*+t)},&\cal{H}^-_{\textit{r}_-}
    \end{cases}.
\end{equation}
These modes are shown with solid arrows in Fig.~\ref{fig:RN_penrose}. They represent the experience of any inertial observer with arbitrary energy $E_\text{ob}$ (up to a rescaling of the frequency $\omega$); without loss of generality, an observer with ${E_\text{ob}=1}$ is chosen for the left potion of the inner horizon in Fig.~\ref{fig:RN_penrose} ($\cal{H}^-_{\textit{r}_-}$), while an observer with ${E_\text{ob}=-1}$ is chosen for the right portion ($\cal{H}^-_{\textit{r}_-}$). Also, note that if the observer is placed at the outer horizon instead of the inner horizon, a similar complete set of modes can be defined \textit{mutatis mutandis}. In what follows, we will present the results for both sets of modes simultaneously, though we will only closely follow the steps of analysis for the inner horizon observers' set of modes.

Equipped with a complete set of modes for an Unruh emitter and an inertial observer, one may now proceed to calculate the expectation value of the particle number operator seen by the observer in the emitter's vacuum state via Eq.~(\ref{eq:bog}). To do so, consider what will subsequently be referred to as the past null Cauchy hypersurface, consisting of the union of past null infinity with the exterior and interior past horizons ($\mathscr{I}^-\cup\ \cal{H}^+_{\textit{r}_+}$; see Fig.~\ref{fig:RN_penrose}). On this surface, the emitter's modes are given by Eqs.~(\ref{eq:RN_f_em+}) and (\ref{eq:RN_f_em-}), while the observer's modes can be found with scattering theory, as described below.

Since the $t$ coordinate used to define the observer's modes defines a global timelike Killing vector for the spacetime, the field's modes $f_{\omega\ell}$ can be separated as
\begin{equation}\label{eq:RN_f_chi}
    f_{\omega\ell}(t,r^*)\equiv\chi_{\omega\ell}(r^*)\ \text{e}^{\pm i\omega t}.
\end{equation}
This separation puts the Klein-Gordon wave Eq.~(\ref{eq:waveeq_RN_tr}) into the form of a 1D scattering equation in $r^*$. In the limits as $\Delta$ approaches both 0 and 1, the scattering potential of Eq.~(\ref{eq:waveeq_RN_tr}) vanishes, leading to asymptotic eigenmode solutions of the form ${\exp(\pm i\omega r^*)}$. As such, the observer's modes $\chi^\pm_\text{ob}$ can be backpropagated to the past null Cauchy hypersurface\textemdash altogether, for an observer at future null infinity one has
\begin{equation}\label{eq:RN_fob+ext}
    {}^\text{ext}f^+_\text{ob}\to\text{e}^{-i\omega t}
    \begin{cases}
        \text{e}^{i\omega r^*}+\mathcal{R}^+_\text{ext}\text{e}^{-i\omega r^*},&r^*_\text{ext}\to\infty\\
        \mathcal{T}^+_\text{ext}\text{e}^{i\omega r^*},&r^*_\text{ext}\to-\infty
    \end{cases},
\end{equation}
for an outgoing observer at the inner horizon,
\begin{equation}\label{eq:RN_fob+int}
    {}^\text{int}f^+_\text{ob}\to\text{e}^{i\omega t}
    \begin{cases}
        \text{e}^{-i\omega r^*},&r^*_\text{int}\to\infty\\
        \mathcal{T}^+_\text{int}\text{e}^{-i\omega r^*}+\mathcal{R}^+_\text{int}\text{e}^{i\omega r^*},&r^*_\text{int}\to-\infty\\
        \mathcal{R}^+_\text{int}\mathcal{T}^-_\text{ext}\text{e}^{i\omega r^*},&r^*_\text{ext}\to\infty\\
        \mathcal{R}^+_\text{int}\left(\text{e}^{i\omega r^*}+\mathcal{R}^-_\text{ext}\text{e}^{-i\omega r^*}\right),&r^*_\text{ext}\to-\infty
    \end{cases},
\end{equation}
and for an ingoing observer at the inner horizon,
\begin{equation}\label{eq:RN_fob-int}
    {}^\text{int}f^-_\text{ob}\to\text{e}^{-i\omega t}
    \begin{cases}
        \text{e}^{-i\omega r^*},&r^*_\text{int}\to\infty\\
        \mathcal{T}^-_\text{int}\text{e}^{-i\omega r^*}+\mathcal{R}^-_\text{int}\text{e}^{i\omega r^*},&r^*_\text{int}\to-\infty\\
        \mathcal{T}^-_\text{int}\mathcal{T}^-_\text{ext}\text{e}^{-i\omega r^*},&r^*_\text{ext}\to\infty\\
        \mathcal{T}^-_\text{int}\left(\text{e}^{-i\omega r^*}+\mathcal{R}^-_\text{ext}\text{e}^{i\omega r^*}\right),&r^*_\text{ext}\to-\infty
    \end{cases},
\end{equation}
where $r^*_\text{int}$ and $r^*_\text{ext}$ represent the radial tortoise coordinates $r^*$ for the black hole's interior and exterior, respectively. The reflection coefficients $\mathcal{R}^\pm_{\text{int,ext}}$ and transmission coefficients $\mathcal{T}^\pm_{\text{int,ext}}$, which depend on the observer's mode numbers $\omega$ and $\ell$, can be computed numerically (or semianalytically with confluent Heun functions) with the above boundary conditions on the wave Eq.~(\ref{eq:waveeq_RN_tr}); see the \hyperref[sec:appendix]{Appendix} for more details.

Defining annihilation operators ${{}^{\text{int,ext}}a^\pm_\text{ob,em}}$ for each respective set of modes ${{}^{\text{int,ext}}f^\pm_\text{ob,em}}$, we can now calculate the particle content seen by the observer. The vacuum expectation values of the number operators associated with each choice of observer are
\begin{equation}
    \langle N^\pm_{\text{int,ext}}\rangle\equiv\left\langle0_\text{em}\left|\left({}^{\text{int,ext}}a^\pm_\text{ob}\right)^\dagger\left({}^{\text{int,ext}}a^\pm_\text{ob}\right)\right|0_\text{em}\right\rangle,
\end{equation}
with ${\langle N^+_\text{ext}\rangle}$ for the expected particle number observed at future null infinity $\mathscr{I}^+$, ${\langle N^-_\text{ext}\rangle}$ for an observer at the event horizon $\cal{H}^-_{\textit{r}_+}$, ${\langle N^+_\text{int}\rangle}$ for an outgoing observer at the inner horizon $\cal{H}^+_{\textit{r}_-}$, and ${\langle N^-_\text{int}\rangle}$ for an ingoing observer at the inner horizon $\cal{H}^-_{\textit{r}_-}$.

Using Eqs.~(\ref{eq:bog}) and (\ref{eq:inner_product}) and evaluating the inner product between the emitter's modes and the observer's backpropagated modes along the past null Cauchy hypersurface, the anticipated number operators can be calculated. After summing over the angular modes, the following inner products yield nontrivial (i.e.\ up to an irrelevant phase) contributions to the Bogoliubov coefficients:
\begin{subequations}\label{eq:Nschem}
\begin{align}
    \langle N^+_\text{ext}\rangle&=\int_0^\infty d\bar{\omega}\ \left|\Tpext\left\langle\text{e}^{-i\bar{\omega}U}\big|\text{e}^{i\omega u}\right\rangle_{{}^\text{ext}\cal{H}^+_{\textit{r}_+}}\right|^2,\\\label{eq:Nextm}
    \langle N^-_\text{ext}\rangle&=\int_0^\infty d\bar{\omega}\ \left|\Rmext\left\langle\text{e}^{-i\bar{\omega}U}\big|\text{e}^{i\omega u}\right\rangle_{{}^\text{ext}\cal{H}^+_{\textit{r}_+}}\right|^2,\displaybreak[0]\\
    \langle N^-_\text{int}\rangle&=\int_0^\infty d\bar{\omega}\ \bigg|\RmextTmint\left\langle\text{e}^{-i\bar{\omega}U}\big|\text{e}^{i\omega u}\right\rangle_{{}^\text{ext}\cal{H}^+_{\textit{r}_+}}\nonumber\\
    &\hspace{4em}+\Rmint\left\langle\text{e}^{-i\bar{\omega}U}\big|\text{e}^{i\omega u}\right\rangle_{{}^\text{int}\cal{H}^+_{\textit{r}_+}}\bigg|^2,\displaybreak[0]\\
    \langle N^+_\text{int}\rangle&=\int_0^\infty d\bar{\omega}\ \bigg|\Tpint\left\langle\text{e}^{-i\bar{\omega}U}\big|\text{e}^{-i\omega u}\right\rangle_{{}^\text{int}\cal{H}^+_{\textit{r}_+}}\nonumber\\
    &\hspace{4em}+\RmextRpint\left\langle\text{e}^{-i\bar{\omega}U}\big|\text{e}^{-i\omega u}\right\rangle_{{}^\text{ext}\cal{H}^+_{\textit{r}_+}}\bigg|^2\nonumber\\
    &+\int_0^\infty d\bar{\omega}\ \bigg|\TmextRpint\left\langle\text{e}^{-i\bar{\omega}v}\big|\text{e}^{-i\omega v}\right\rangle_{\mathscr{I}^-}\bigg|^2,\label{eq:Nintp}
\end{align}
\end{subequations}
where each Penrose diagram stands in for the complex conjugate of the backscattering coefficient(s) corresponding to the path shown, e.g.\ the final diagram of Eq.~(\ref{eq:Nintp}) represents the combination ${(\mathcal{R}^+_\text{int}\mathcal{T}^-_\text{ext})^*}$, and the subscript associated with each bra-ket indicates the null surface over which that inner product is evaluated.

Several potential pathways appear to be missing from Eqs.~(\ref{eq:Nschem}), such as the pathway in Eq.~(\ref{eq:Nextm}) connecting the event horizon to past null infinity. However, all such pathways involve inner products of the form ${\langle\text{e}^{-i\bar{\omega}v}|\text{e}^{i\omega v}\rangle}$, whose modes are completely orthogonal and therefore do not contribute at all to the Bogoliubov coefficients. While the exclusion of these pathways is entirely straightforward and routine, one may wonder how these calculations relate to those of Sec.~\ref{sec:redshift}, where, for example, an observer at the event horizon \emph{does} see nontrivial contributions from the ingoing modes from past null infinity. The discrepancy lies in the fact that adiabaticity is never satisfied for $\kappa^-(r_+)$, and therefore the effective temperature calculations cannot be trusted at that specific location. There is a small range of black hole charges around the value ${Q/M\approx0.937}$ for which it appears from Fig.~\ref{fig:epsilon_RN} that $\epsilon$ dips below 1 at the event horizon; however, as noted in Sec.~\ref{subsec:adi}, in this range, the adiabatic control function $\epsilon$ fails to be a good estimator of the degree of adiabaticity, since higher derivatives of $\kappa$ dominate over the vanishing first derivative.

The evaluation of the inner products from Eqs.~(\ref{eq:Nschem}) over each relevant surface have become a standard part of the literature (see, e.g., Refs.~\cite{bar11a}, \cite{bal19}, and sources therein); for example,
\begin{equation}
    \left|\left\langle\text{e}^{-i\bar{\omega}U}\big|\text{e}^{i\omega u}\right\rangle_{{}^\text{ext}\cal{H}^+_{\textit{r}_+}}\right|^2=\frac{\text{e}^{-\pi\omega/\varkappa_+}}{4\pi^2\varkappa_+^2}\frac{\omega}{\bar{\omega}}\left|\Gamma\left(\frac{i\omega}{\varkappa_+}\right)\right|^2
\end{equation}
Using the property of gamma functions
\begin{equation}
    \left|\Gamma(\pm ix)\right|^2=\frac{2\pi}{x\left(\text{e}^{\pi x}-\text{e}^{-\pi x}\right)},
\end{equation}
one finds a Planckian distribution in the observer's frequency $\omega$, and while the remaining factors of ${1/(2\pi\varkappa_+\bar{\omega})}$ formally diverge when the integrals of Eqs.~(\ref{eq:Nschem}) are carried out, this divergence only occurs as a result of the unphysical usage of infinite plane waves. If one were instead to use a normalized wave packet localized in each asymptotic region with a frequency content concentrated around some frequency $\bar{\omega}^*$, the offending terms would all reduce to unity.

Since the scattering coefficients in Eqs.~(\ref{eq:Nschem}) are independent of the emitter's modes $\bar{\omega}$ (the emitter's modes are kept at their initial past boundaries, while the observer's modes are the ones that must be backpropagated through the spacetime's scattering potential), the final form of the number operators at each surface simplifies to (cf.~the number operators of Ref.~\cite{bal19} evaluated for a simplified scattering potential):
\begin{subequations}\label{eq:N}
\begin{align}
    \langle N^+_\text{ext}\rangle&=\frac{\left|\mathcal{T}^+_\text{ext}\right|^2}{\text{e}^{2\pi\omega/\varkappa_+}-1},\displaybreak[0]\\
    \langle N^-_\text{ext}\rangle&=\frac{\left|\mathcal{R}^-_\text{ext}\right|^2}{\text{e}^{2\pi\omega/\varkappa_+}-1},\displaybreak[0]\\
    \langle N^-_\text{int}\rangle&=\frac{\left|\mathcal{T}^-_\text{int}\mathcal{R}^-_\text{ext}-\mathcal{R}^-_\text{int}\text{e}^{\pi\omega/\varkappa_+}\right|^2}{\text{e}^{2\pi\omega/\varkappa_+}-1},\\
    \langle N^+_\text{int}\rangle&=\frac{\left|\mathcal{T}^+_\text{int}-\mathcal{R}^+_\text{int}\mathcal{R}^-_\text{ext}\text{e}^{\pi\omega/\varkappa_+}\right|^2}{\text{e}^{2\pi\omega/\varkappa_+}-1}+\left|\mathcal{R}^+_\text{int}\mathcal{T}^-_\text{ext}\right|^2.
\end{align}
\end{subequations}

The key feature in each of the above equations is the familiar Planckian spectrum ${(\text{e}^{2\pi\omega/{\varkappa_+}}-1)^{-1}}$, modified by a frequency-dependent graybody factor associated with the appropriate set of scattering coefficients. For example, if no modes were scattered in the black hole exterior (and therefore ${\mathcal{T}^+_\text{ext}=1}$), Eq.~(\ref{eq:N}a) would reduce to a completely thermal Hawking spectrum, as expected for an eikonal observer at infinity.

\subsection{Results}

\begin{figure*}[t]
\centering
\begin{minipage}[l]{0.85\columnwidth}
  \includegraphics[width=\columnwidth]{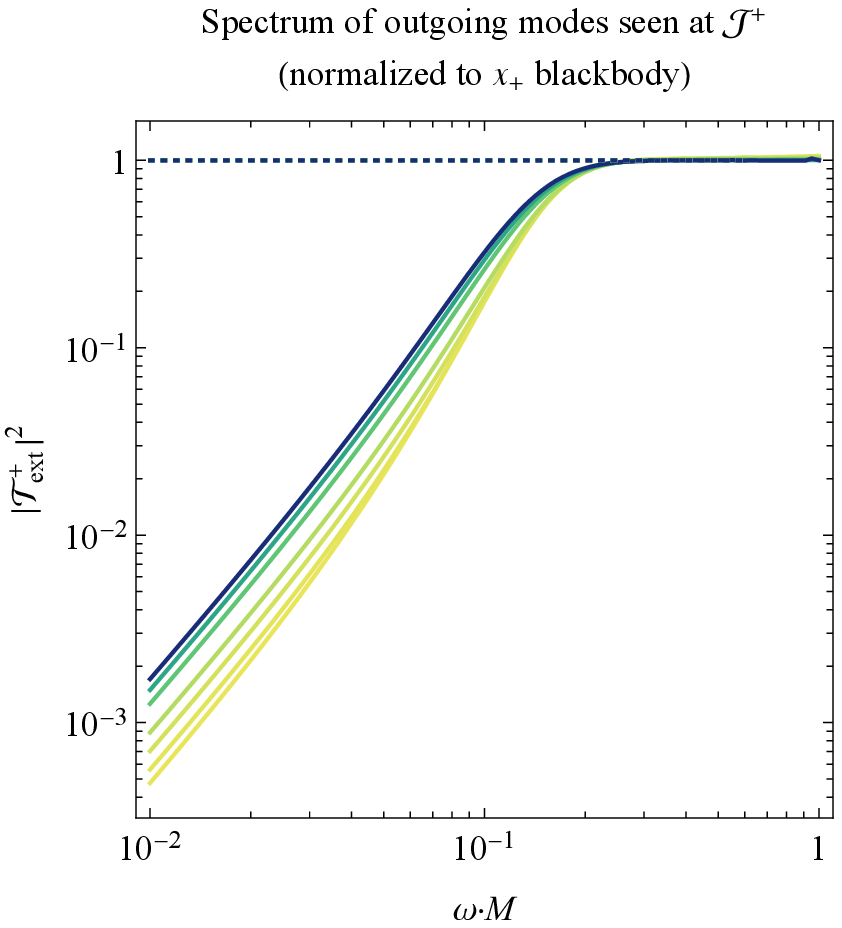}
\end{minipage}%
\hspace{1.5em}
\begin{minipage}[r]{0.85\columnwidth}
    \includegraphics[width=\columnwidth]{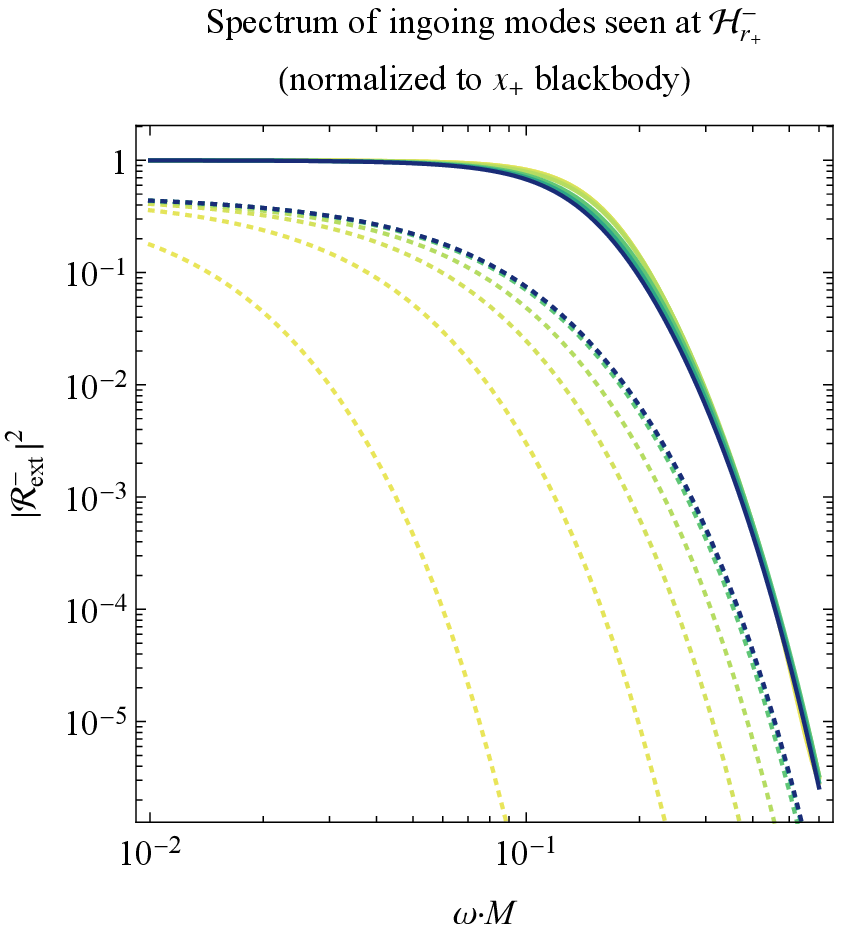}
\end{minipage}%
\vspace{1em}
\begin{minipage}[l]{0.85\columnwidth}
  \includegraphics[width=\columnwidth]{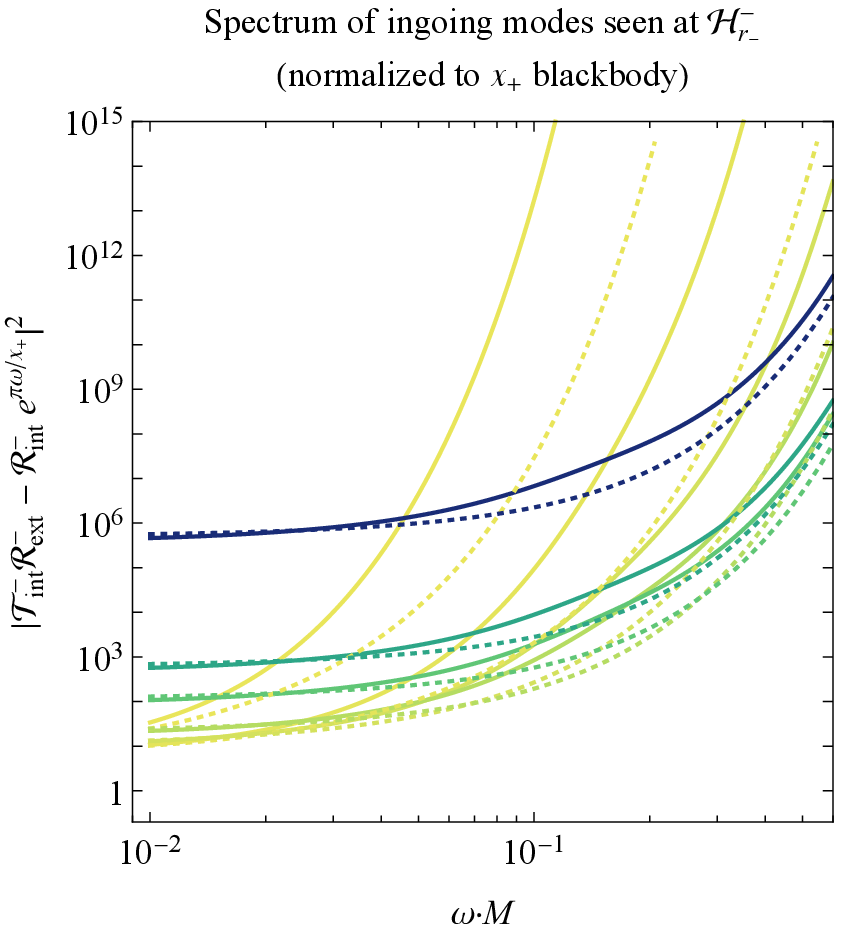}
\end{minipage}%
\hspace{1.5em}
\begin{minipage}[r]{0.85\columnwidth}
    \includegraphics[width=\columnwidth]{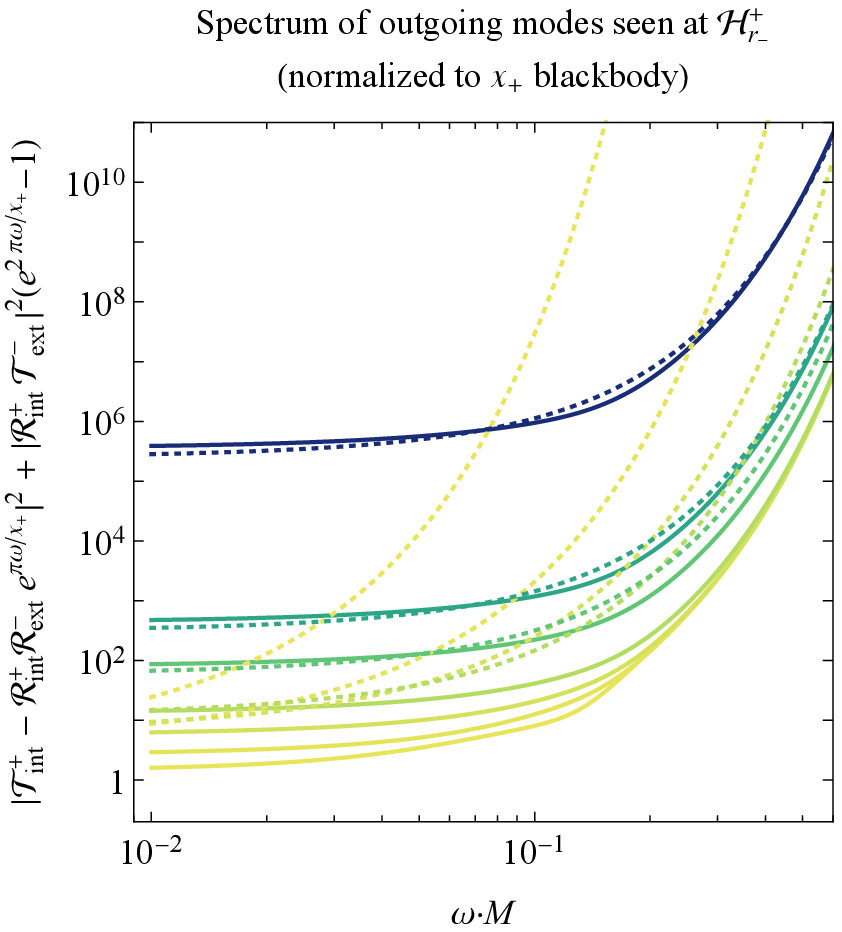}
\end{minipage}
\caption{Graybody $s$-mode factors from Eqs.~(\ref{eq:N}) modifying the thermal ${\varkappa_+/(2\pi)}$-temperature Hawking spectra seen by an observer asymptotically far from the black hole looking downward at outgoing modes (top left), an ingoing observer at the event horizon looking upward at ingoing modes (top right), an ingoing observer at the inner horizon looking upward at ingoing modes (bottom left), and an outgoing observer at the inner horizon looking downward at outgoing modes (bottom right). Different black hole charges are shown with respective colors from dark blue to yellow: ${Q/M}$ = 0.1, 0.5, 0.7, 0.9, 0.96, 0.99, and 0.999. Solid curves show the numerically computed spectra, while dotted curves show the positive-valued spectra obtained from a completely thermal distribution with temperatures $\kappa^+/(2\pi)$ from Eq.~(\ref{eq:RN_kappa_inf}) (upper left), $\kappa^-/(2\pi)$ from Eq.~(\ref{eq:RN_kappa_r+}) (upper right), $\kappa^-/(2\pi)$ from Eq.~(\ref{eq:RN_kappa_r-}) (lower left), or $\kappa^+/(2\pi)$ from Eq.~(\ref{eq:RN_kappa_r-_E-1}) (lower right).\label{fig:spectra}}
\end{figure*}

\subsubsection{Spectra for s-modes}

Fig.~\ref{fig:spectra} shows the deviations from thermality for the ${\ell=0}$ spectra of Eqs.~(\ref{eq:N}). These plots are computed numerically with the help of confluent Heun functions, as outlined in the \hyperref[sec:appendix]{Appendix}. In the top left panel, the particle spectrum seen asymptotically far from the black hole is plotted as the ratio of ${\langle N^+_\text{ext}\rangle}$ to the analogous occupation number for a fully thermal spectrum with temperature ${\varkappa_+/(2\pi)}$ (this Planckian distribution will subsequently be referred to as a ``$\varkappa_+$ blackbody''). This ratio, which from Eq.~(\ref{eq:N}a) equals the transmission probability ${\left|\mathcal{T}^+_\text{ext}\right|^2}$, approaches unity in the high frequency (geometric optics) limit, indicating a return to thermality in that regime. However, at lower frequencies, significant deviation from thermality occurs as the spectrum attains a steeper power law than that of a blackbody. The transmission probability approaches a power law index of 2, as first predicted by Starobinsky and Churilov for the analytically solvable ${\omega M\ll1}$ regime \cite{sta73}.

For an observer crossing the event horizon, the Hawking radiation seen from ingoing modes in the sky above is shown in the top right panel of Fig.~\ref{fig:spectra}. Just as in the top left panel, values at unity indicate consistency with a $\varkappa_+$ blackbody spectrum, though in this case, thermality at the surface gravity temperature mostly occurs at the lowest frequencies instead of the highest frequencies, with slight deviations for different black hole charges $Q$. At higher frequencies, the spectrum cuts off much sooner than it does for an asymptotically infinite observer, indicating a lower eikonal temperature. This high-frequency-limit temperature (multiplied by $2\pi$) is approximately, but not exactly, equal to the effective temperature $\kappa^-$ from Eq.~(\ref{eq:RN_kappa_r+}), as shown by the dotted curves in Fig.~\ref{fig:spectra}. Indeed, the modes contributing to the Bogoliubov spectrum from Eq.~(\ref{eq:Nextm}) are ingoing at the event horizon and therefore tied to $\kappa^-$, although adiabaticity is not quite satisfied there.

In principle, one may also calculate the spectrum of \emph{outgoing} Hawking modes seen at the event horizon, corresponding to the effective temperature $\kappa^+$ there, and indeed, an infalling observer will still see an exponentially redshifting past horizon below them after they cross the event horizon. However, calculating the outgoing modes for an ingoing horizon observer (and vice versa) requires Fourier-decomposing the observer's modes of Eq.~(\ref{eq:U(u)}) so that they can be backpropagated to the past horizon, which will be deferred to a future study; for more details, see, e.g., Ref.~\cite{lan18}.

Though only frequencies as high as ${\omega M\sim0.6}$ are shown for the horizon spectra of Fig.~\ref{fig:spectra} (the ${\omega M\gg1}$ regime is beyond our current numerical capabilities), any higher frequencies are all but irrelevant compared to the luminosity peaks of the blackbodies, which, though not shown in the normalized spectra of Fig.~\ref{fig:spectra}, occur between ${\omega M\sim0.2}$ (for the lowest charge $Q$) and ${\omega M\sim0.01}$ (for the highest charge plotted).

While the Hawking spectra seen at infinity and the event horizon contain straightforward graybody deviations from a thermal spectrum, the spectra seen at the inner horizon tell a different story. Two spectra for the left and right portions of the inner horizon are shown in the lower left and right panels of Fig.~\ref{fig:spectra}, respectively. These spectra bear little resemblance to the initial $\varkappa_+$ blackbodies seen at infinity; nonetheless, we still present the spectra normalized to the $\varkappa_+$ blackbodies due to the factors in the denominators of Eqs.~(\ref{eq:N}).

At the left leg of the inner horizon (lower left panel of Fig.~\ref{fig:spectra}), the particle spectrum given by the Bogoliubov coefficient between the observer's and emitter's vacuum states all appear to be ultraviolet-divergent; if an exponential cutoff does occur, it must happen at frequencies higher than we are able to calculate. A qualitatively similar spectrum would occur for a Planckian distribution with negative temperature (albeit with an overall sign change), as anticipated in Secs.~\ref{subsec:rad} and \ref{subsec:ang}, and for reference, the corresponding negative-temperature $\kappa^-$ blackbodies are shown by the dotted curves in Fig.~\ref{fig:spectra}. Notably, as ${Q/M\to1}$, the ultraviolet divergence grows stronger, though as ${Q/M\to0}$, the entire spectrum diverges (once $Q/M$ goes below $\sim0.01$, the spectrum is too high to be seen on these lower two plots). Such a panchromatic divergence can be attributed to the fact that the inner horizon's surface gravity $\varkappa_-$, and consequently the temperatures $\kappa^-$ from Eq.~(\ref{eq:RN_kappa_r-}) and $\kappa^+$ from Eq.~(\ref{eq:RN_kappa_r-_E-1}), grow to infinity in the Schwarzschild limit, since $r_-\to0$.

At the inner horizon's right leg (lower right panel of Fig.~\ref{fig:spectra}), the curves once again diverge at higher frequencies, indicating quasitemperatures much higher than the underlying $\varkappa_+$ blackbodies. These temperatures may be high enough to be negative, though when the black hole charge is large enough, the spectra begin to deviate significantly from the dotted lines showing $\kappa^-$ blackbodies. Nonetheless, the spectrum is still everywhere nonthermal as a result of the frequency-dependent additive final term in Eq.~(\ref{eq:N}d).

\subsubsection{Spectra for higher spherical harmonics}

\begin{figure}[t!]
  \includegraphics[width=0.9\columnwidth]{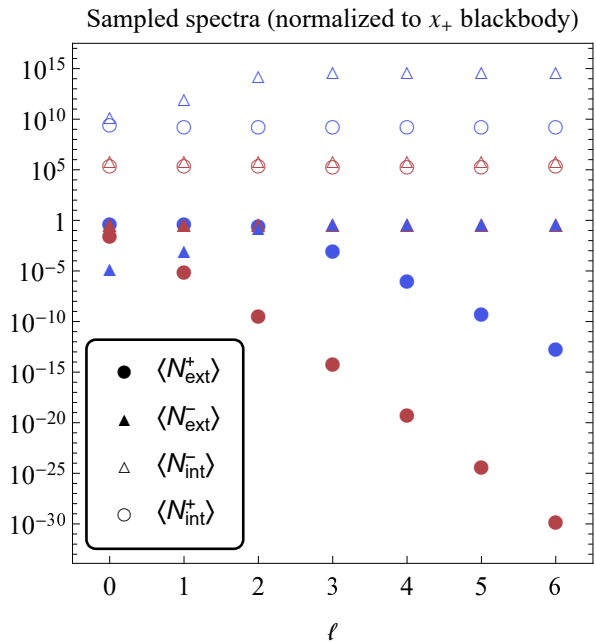}
  \caption{Sampled points at ${\omega M=0.5}$ (blue) and ${\omega M=0.05}$ (red) for the four spectra of Fig.~\ref{fig:spectra} when generalized to higher-$\ell$ modes. All points use a black hole charge of ${Q/M=0.1}$. The ${\ell=0}$ mode dominates the spectrum ${\langle N^+_\text{ext}\rangle}$ seen at infinity, but higher-$\ell$ modes make substantial contributions to the spectra seen at the horizons.\label{fig:Nl}}
\end{figure}

The dependence of the Hawking spectra on the spherical harmonic mode number $\ell$ is shown in Fig.~\ref{fig:Nl}. Instead of plotting the entire spectrum for each $\ell$, we sample two points from each spectrum, one at a higher frequency (${\omega M=0.5}$, blue points) and one at a lower frequency (${\omega M=0.05}$, red points). In almost all cases (except the spectra for ${\langle N^-_\text{ext}\rangle}$; see the upper right panel of Fig.~\ref{fig:spectra}), the higher-frequency blue points exceed their lower-frequency red counterparts, indicating that the general qualitative trends of each spectrum in Fig.~\ref{fig:spectra} remain intact for higher-$\ell$ modes.

For the Hawking radiation seen asymptotically far from the black hole, the ${\ell=0}$ mode dominates over any higher harmonics \cite{pag76}, as can be seen from the drop-off of the solid circular points in Fig.~\ref{fig:Nl}. However, for radiation seen at the outer and inner horizons, the spectra do not seem to fall off as $\ell$ is increased. It would appear that the ultraviolet-divergent Hawking spectra contain substantial contributions not only from the spherical ${\ell=0}$ modes, but also from much higher harmonics. One important implication of this result is that semiclassical calculations of the renormalized stress-energy tensor in the (1+1)D Polyakov approximation potentially miss out on key beyond-$s$-wave physics near the horizons.

\section{\label{sec:dis}Discussion}

Two of the main questions underpinning this study are as follows: how would Hawking radiation appear for someone at a black hole's inner horizon? And what is meant by a negative Hawking temperature in this context? Ultimately, one may wish to understand the full quantum backreaction near the inner horizon, and though we are not at a place to provide a definitive assertion regarding the fully dynamical, quantum gravitational backreaction, the present analysis does shed further light on the nature of both Hawking radiation and semiclassical charged black holes.

To study the Hawking radiation seen anywhere near or far from a black hole, we began with the effective temperature functions $\kappa^\pm$ for an observer looking radially inward or outward \cite{bar11a,bar11b}, as defined in Sec.~\ref{subsec:foreff}, which reproduces Hawking's original calculation in the geometric optics limit but generalizes to an arbitrary inertial observer at a radius $r_\text{ob}$. This effective temperature, given by Eqs.~(\ref{eq:RN_kappa}) for an infaller from rest at infinity, diverges at the inner horizon, and regardless of the observer's orbital parameters, it becomes negative (indicative of modes that are blueshifting instead of redshifting) once the observer falls close enough to the inner horizon. As it turns out, this negative temperature is not merely confined to the black hole's interior that would remain inaccessible to the outside universe; instead, when the charge-to-mass ratio is high enough, specifically when ${(Q/M)^2>8/9}$, the inner horizon becomes close enough to the event horizon that a negative $\kappa^+$ is detected \emph{outside} the black hole.

The change in sign of the effective Hawking temperature for observers close enough to the inner horizon was found in Sec.~\ref{subsec:ang} to occur not just in the radial direction, but in every direction the observer looks in their field of view. The classical phenomenon of mass inflation involves a divergence only at a single radial point in the sky (as an outgoing observer approaches the inner horizon, the sky above them will shrink to a point and become infinitely blueshifted), but semiclassically, Hawking radiation originating from the past horizon will fill the observer's entire field of view with diverging, negative-temperature radiation as they approach the inner horizon.

Are the approximations of the effective temperature formalism even valid whenever $\kappa$ becomes negative? By studying the adiabatic control function $\epsilon$ in Sec.~\ref{subsec:adi}, we can learn whether $\epsilon$ is small enough for the adiabatic condition to be satisfied and therefore for $\kappa$ to reproduce approximately thermal Bogoliubov coefficients. We find that at the inner horizon, the outgoing modes for an ingoing observer are sufficiently adiabatic for a large enough black hole charge $Q$, while the ingoing modes are never adiabatic there.

To complement these effective temperature results and provide a more rigorous calculation in the regimes where the adiabatic condition fails, we finally performed a full wave mode analysis in Sec.~\ref{sec:spe} to determine the Bogoliubov spectrum at each of the asymptotic regimes. To do so, the observer's wave modes were backpropagated through the spacetime to the position of the Unruh emitter using Eq.~(\ref{eq:waveeq_RN_tr}) for a massless scalar Klein-Gordon field, and the inner product of the observer's and emitter's modes was computed. Asymptotically far from the black hole, the spectrum becomes completely thermal for high enough frequencies (i.e.\ in the geometric optics limit), which is consistent with the vanishing of the outgoing adiabatic control function $\epsilon^+$ at infinity. In contrast, for an observer at the event horizon, $\epsilon^+$ is almost never significantly smaller than unity, and the corresponding Bogoliubov spectrum does deviate significantly from thermality in the geometric optics limit.

At the inner horizon, the spectrum of scalar particles appears quite different from that of a positive-temperature blackbody, and instead looks much more like the spectrum one would obtain (up to an overall change in sign) from a blackbody with a negative temperature. The spectra are thus mostly consistent with the effective temperature predictions, despite the general lack of adiabaticity in that regime. The familiar Rayleigh-Jeans power law is still present at lower frequencies, but at higher frequencies, the spectral intensity continues to climb even higher. Even if an exponential decay at higher frequencies never occurs with the present formalism, one may nonetheless suspect that some ultraviolet cutoff will exist once the semiclassical approximation breaks down at the Planck scale, or, more importantly, that the semiclassical backreaction in a dynamical collapse would prevent such a spectrum from ever occurring in the first place.

The right leg of the inner horizon is unique in that its spectrum contains substantial contributions not only from the outgoing Unruh modes originating from the past horizon below the observer (as in all the other cases), but also from the ingoing Unruh modes originating from the sky above; see Eq.~(\ref{eq:Nintp}). The resulting spectra are even more divergent at low $Q$ than those of the left leg of the inner horizon, though at higher $Q$, the spectra appear much tamer (albeit still with a much higher graybody temperature than that observed asymptotically far away). If one wishes to compare these spectra with the effective temperature formalism of Sec.~\ref{sec:redshift}, it should be noted that the adiabatic control functions of Fig.~\ref{fig:epsilon_RN} are only valid for the left leg of the inner horizon\textemdash for the right leg, $\epsilon^-$ always equals 1, while $\epsilon^+$ is always greater than 1.

When comparing the inner horizon values of the effective temperatures $\kappa^\pm$ with their corresponding Bogoliubov spectra, it is important to note that the two spectra shown in the lower panels of Fig.~\ref{fig:spectra} are associated with the Hawking sectors that are \emph{not} expected to yield diverging effective temperatures (but are nonetheless negative); namely, the ingoing temperature $\kappa^-$ in Eq.~(\ref{eq:RN_kappa_r-}) and the outgoing temperature $\kappa^+$ in Eq.~(\ref{eq:RN_kappa_r-_E-1}). If an ingoing (or outgoing) observer at the inner horizon looks downward (or upward, respectively), they should be met with an even stronger dose of diverging Hawking radiation. But what Fig.~\ref{fig:spectra} communicates is that for an outgoing observer approaching the inner horizon, while they can look upward to see the Penrose blueshift singularity forming, if they look downward at the initially dimming and redshifting past horizon, even this surface will eventually begin to blueshift and produce a semiclassically divergent spectrum of Hawking radiation.

The implications of these Hawking spectra are clear: the interaction of a quantum scalar field with a charged black hole results in runaway particle creation detected at the inner horizon. The particle spectrum diverges at all frequencies as ${Q/M\to0}$, since the inner horizon coincides with the ${r=0}$ singularity that was already found to feature a diverging Hawking flux in Ref.~\cite{ham18}. But even for nonzero charge, the inner horizon spectrum becomes highly blueshifted and is potentially ultraviolet-divergent.

Such a highly energetic source of radiation will quickly become amplified in the radial direction and provide an ongoing source for the Poisson-Israel mass inflation instability. Even if the observer is taken to be something as simple as a two-level atom, one may speculate that the implied Hawking flux would energize the atom to such an extent that the inevitable result is a positive feedback loop resulting in the collapse of the spacetime geometry into a spacelike singularity. The Reissner-Nordstr\"om metric cannot remain semiclassically intact; its inner horizon must collapse into a singularity or else evolve dynamically into some potentially horizonless object.

Several important questions remain to be answered. While the effective temperature and Bogoliubov spectrum formalisms complement each other well in many regards, the exact conditions under which they agree remain to be proven, especially since the adiabatic control function $\epsilon$ often fails to predict when the spectrum will or will not appear Planckian at high frequencies (or even whether $\kappa$ itself is adiabatic). Additionally, one may wish to explore further the implications of the dominance of higher-$\ell$ Hawking modes once an observer reaches either horizon (note that the higher-$\ell$ modes of Fig.~\ref{fig:Nl} cannot be directly compared with the observed angular modes of Fig.~\ref{fig:kappa_chi}, since the latter describe angular modes with respect to the observer while the former describe angular modes with respect to the black hole's center). But the biggest question one may wish to ask is whether either calculation is able to predict the presence of ``real'' particles. We have not made use of any response functions, Unruh-DeWitt detectors, or renormalization schemes that would indicate the influence of a Hawking particle on an observer or on the underlying spacetime geometry. Nonetheless, in analyzing how the effective temperature depends on an observer's energy, it does appear to preserve Lorentz covariance in some regimes, and regardless, there is no doubt that the semiclassical effects predicted here should substantially alter the spacetime geometry near the inner horizon.

\appendix*
\section{\label{sec:appendix}Backscattering coefficients via confluent Heun functions}

In this appendix we outline the methodology to compute the backscattering coefficients used in Sec.~\ref{sec:spe} to find the graybody factors associated with the Hawking spectrum at infinity, the event horizon, and the inner horizon. Eqs.~(\ref{eq:RN_fob+ext})$-$(\ref{eq:RN_fob-int}) provide the boundary conditions for the observer's backscattered mode functions in terms of the reflection coefficients $\mathcal{R}^{\pm}_{\text{int,ext}}$ and transmission coefficients $\mathcal{T}^{\pm}_{\text{int,ext}}$, where the subscript labels whether the scattering occurs in the black hole's interior (``int'') or exterior (``ext''), and the superscript labels whether the modes are outgoing ($+$) or ingoing ($-$) prior to backpropagation, at the future null surface in the relevant spacetime sector. Conservation of the Wronskian dictates that these coefficients satisfy the following normalization conditions:
\begin{subequations}\label{eq:norm}
\begin{align}
    \left|\mathcal{T}^\pm_\text{int}\right|^2-\left|\mathcal{R}^\pm_\text{int}\right|^2&=1,\\
    \left|\mathcal{T}^\pm_\text{ext}\right|^2+\left|\mathcal{R}^\pm_\text{ext}\right|^2&=1,
\end{align}
\end{subequations}
which will provide a check to ensure the accuracy of the numerical scheme. The negative sign associated with $\mathcal{R}^\pm_\text{int}$ in Eq.~(\ref{eq:norm}a) is due to the fact that the corresponding substates have a negative norm; the scattering potential inside the black hole allows for the existence of both the observer's original modes ${\exp(-i\omega r^*)}$ (positive frequency with respect to the timelike coordinate $r^*$) and the anomalous modes ${\exp(+i\omega r^*)}$.

The backscattering coefficients can be calculated either by implementing an implicit numerical ODE method to solve the Klein-Gordon wave Eq.~(\ref{eq:waveeq_RN_tr}), or by matching analytic solutions to that equation. Here we will explore the latter option.

Instead of the mode separation of Eq.~(\ref{eq:RN_modesep}), the Klein-Gordon scalar field can be separated as
\begin{equation}
    \phi_{\omega\ell m}=\frac{R_{\omega\ell}(r)\ \text{e}^{\pm i\omega t}\ Y_{\ell m}(\theta,\varphi)}{\sqrt{4\pi\omega}},
\end{equation}
with the upper ($+$) sign in the exponential for the outgoing modes observed at the right leg of the inner horizon (which can be written as ${{}^\text{int}R^+_\text{ob}}$) and the lower ($-$) sign for both the ingoing modes observed at the left leg of the inner horizon (${{}^\text{int}R^-_\text{ob}}$) as well as the outgoing modes observed at future null infinity (${{}^\text{ext}R^+_\text{ob}}$). In terms of the modes of Eq.~(\ref{eq:waveeq_RN_tr}), $R_{\omega\ell}$ and $f_{\omega\ell}$ are related by
\begin{equation}
    f_{\omega\ell}(t,r)=rR_{\omega\ell}(r)\ \text{e}^{\pm i\omega t}.
\end{equation}
The Klein-Gordon wave equation for the spatial modes $R_{\omega\ell}$, or equivalently, the wave Eq.~(\ref{eq:waveeq_RN_tr}) for $f_{\omega\ell}$, contains three singular points throughout the spacetime, which occur whenever ${r^*\to\pm\infty}$. Two of these are the regular singularities located at the inner and outer horizons, and the third is an irregular, rank-1 singularity at spatial infinity. This structure suggests that the wave equation can be cast into confluent Heun form: first, apply a M{\"o}bius transformation to define the new coordinate
\begin{equation}\label{eq:z}
    z\equiv\frac{r-r_-}{r_+-r_-}
\end{equation}
so that the singular points are shifted from $r={(r_-,r_+,\infty)}$ to $z={(0,1,\infty)}$. Then, apply a gauge transformation to the field variable that keeps the singular points fixed (such a shift in the Frobenius solution indices is known as an $F$-homotopic transformation):
\begin{equation}\label{eq:F-hom}
    R(z)=z^{\frac{\gamma-1}{2}}|z-1|^{\frac{\delta-1}{2}}\ \text{e}^{\frac{\varepsilon}{2}z}\ Z(z),
\end{equation}
so that the Klein-Gordon wave Eq.~(\ref{eq:waveeq_phi}) reduces to:
\begin{equation}\label{eq:heunc}
    \frac{d^2Z}{dz^2}+\left(\frac{\gamma}{z}+\frac{\delta}{z-1}+\varepsilon\right)\frac{dZ}{dz}+\left(\frac{q}{z}+\frac{\alpha-q}{z-1}\right)Z=0
\end{equation}
provided
\begin{subequations}
\begin{align}
    q&=\ell(\ell+1)+2i\omega\frac{r_+r_-}{r_+-r_-}-4\omega^2r_-^2\nonumber\\
    &+4\omega^2\left(\frac{r_+r_-}{r_+-r_-}\right)^2,\\
    \alpha&=-2i\omega(r_+-r_-)-4\omega^2r_-^2,\displaybreak[0]\\
    \gamma&=1-2i\omega\frac{r_-^2}{r_+-r_-},\\
    \delta&=1-2i\omega\frac{r_+^2}{r_+-r_-},\\
    \varepsilon&=-2i\omega(r_+-r_-).
\end{align}
\end{subequations}
For the more general Kerr-Newman case, the corresponding version of these parameters can be inferred from, e.g.\ Ref.~\cite{vie22}. Also, note that the signs of the three exponents in Eq.~(\ref{eq:F-hom}) can be either positive or negative, corresponding either to outgoing or ingoing waves at each of the singular points. Regardless of this gauge choice, both ingoing and outgoing modes will always be recovered by the choice of linear combinations of modes for $Z(z)$.

Two linearly independent solutions to Eq.~(\ref{eq:heunc}) that are regular at the inner horizon are given via confluent Heun functions for the equation's allowed $F$-homotopic automorphisms:
\begin{subequations}\label{eq:Z_z0}
\begin{align}
    Z_{(0)}(z)&=A_{(0)}Z^A_{(0)}(z)+B_{(0)}Z^B_{(0)}(z),\\
    Z^A_{(0)}(z)&=\text{HeunC}\left(q,\alpha,\gamma,\delta,\varepsilon;z\right),\\
    Z^B_{(0)}(z)&=z^{1-\gamma}\text{HeunC}\left(q',\alpha',2-\gamma,\delta,\varepsilon;z\right),
\end{align}
\end{subequations}
with arbitrary complex coefficients $A_{(0)}$ and $B_{(0)}$, with the definitions
\begin{subequations}
\begin{align}
    q'&=q-(\delta-\varepsilon)(1-\gamma),\\
    \alpha'&=\alpha+\varepsilon(1-\gamma),
\end{align}
\end{subequations}
and with the functions' argument structure following the convention used in \textit{Wolfram Mathematica}, which has newly implemented Heun functions in version 12.1. These negative- and positive-frequency solutions can be computed with a forwardly stable set of power series that are convergent everywhere except at the singular points ${z=1,\infty}$ and are linearly independent except when ${\gamma=1}$, in which case the factor $z^{1-\gamma}$ can be replaced with $\ln(z)$.

As a reminder, the goal here is to compute the values of the reflection and transmission coefficients $\mathcal{R}^\pm_\text{int,ext}$ and $\mathcal{T}^\pm_\text{int,ext}$, which can be used to calculate the observed spectra of Eq.~(\ref{eq:N}). These coefficients are tied to the asymptotic forms of the field modes given in Eqs.~(\ref{eq:RN_fob+ext})$-$(\ref{eq:RN_fob-int}), which in the present notation take the form
\begin{widetext}
\begin{subequations}\label{eq:RN_R_ob}
\begin{align}
    {}^\text{ext}R^+_\text{ob}(z)&\to
    \begin{cases}
        \frac{\text{e}^{i\omega r_-}}{r_+-r_-}\text{e}^{i\omega(r_+-r_-)z}|z|^{2i\omega-1}+\mathcal{R}^+_\text{ext}\frac{\text{e}^{-i\omega r_-}}{r_+-r_-}\text{e}^{-i\omega(r_+-r_-)z}|z|^{-2i\omega-1},&z\to\infty\\
        \mathcal{T}^+_\text{ext}\frac{\text{e}^{i\omega r_+}}{r_+}|z-1|^{i\omega\frac{r_+^2}{r_+-r_-}},&z\to1\\
        0,&z\to0
    \end{cases},\displaybreak[0]\\
    {}^\text{int}R^+_\text{ob}(z)&\to
    \begin{cases}
        \mathcal{R}^+_\text{int}\mathcal{T}^-_\text{ext}\frac{\text{e}^{i\omega r_-}}{r_+-r_-}\text{e}^{i\omega(r_+-r_-)z}|z|^{2i\omega-1},&z\to\infty\\
        \mathcal{R}^+_\text{int}\frac{\text{e}^{i\omega r_+}}{r_+}|z-1|^{i\omega\frac{r_+^2}{r_+-r_-}}+\left(\mathcal{T}^+_\text{int}+\mathcal{R}^+_\text{int}\mathcal{R}^-_\text{ext}\right)\frac{\text{e}^{-i\omega r_+}}{r_+}|z-1|^{-i\omega\frac{r_+^2}{r_+-r_-}},&z\to1\\
        \frac{\text{e}^{-i\omega r_-}}{r_-}|z|^{i\omega\frac{r_-^2}{r_+-r_-}},&z\to0
    \end{cases},\displaybreak[0]\\
    {}^\text{int}R^-_\text{ob}(z)&\to
    \begin{cases}
        \mathcal{T}^-_\text{int}\mathcal{T}^-_\text{ext}\frac{\text{e}^{-i\omega r_-}}{r_+-r_-}\text{e}^{-i\omega(r_+-r_-)z}|z|^{-2i\omega-1},&z\to\infty\\
        \mathcal{T}^-_\text{int}\frac{\text{e}^{-i\omega r_+}}{r_+}|z-1|^{-i\omega\frac{r_+^2}{r_+-r_-}}+\left(\mathcal{R}^-_\text{int}+\mathcal{T}^-_\text{int}\mathcal{R}^-_\text{ext}\right)\frac{\text{e}^{i\omega r_+}}{r_+}|z-1|^{i\omega\frac{r_+^2}{r_+-r_-}},&z\to1\\
        \frac{\text{e}^{-i\omega r_-}}{r_-}|z|^{i\omega\frac{r_-^2}{r_+-r_-}},&z\to0
    \end{cases}.
\end{align}
\end{subequations}
\end{widetext}
%\begin{equation}
%    \text{e}^{i\omega r^*}=\text{e}^{i\omega r_-}\text{e}^{-\frac{\varepsilon}{2}z}\left|z\right|^{-\frac{1-\gamma}{2}}\left|z-1\right|^{\frac{1-\delta}{2}}
%\end{equation}
%\begin{equation}
%    \text{e}^{-i\omega r^*}=\text{e}^{-i\omega r_-}\text{e}^{\frac{\varepsilon}{2}z}\left|z\right|^{\frac{1-\gamma}{2}}\left|z-1\right|^{-\frac{1-\delta}{2}}
%\end{equation}
Here the integration constant for the tortoise coordinate $r^*$ of Eq.~(\ref{eq:tortoise}) is chosen so that
\begin{equation}
    r^*=r+\frac{r_+^2}{r_+-r_-}\ln\left|z-1\right|-\frac{r_-^2}{r_+-r_-}\ln\left|z\right|.
\end{equation}

Asymptotically, the modes of Eq.~(\ref{eq:Z_z0}) at the inner horizon (${z=0}$) reduce to
\begin{equation}
    R(z)\to A_{(0)}z^{-\frac{1-\gamma}{2}}+B_{(0)}z^{\frac{1-\gamma}{2}},\qquad z\to0,
\end{equation}
since the confluent Heun functions are normalized to unity when the independent variable equals zero, provided $\gamma$ is not a nonpositive integer. This asymptotic form can then be matched to the modes of Eq.~(\ref{eq:RN_R_ob}) to find expressions for $A_{(0)}$ and $B_{(0)}$. One obtains ${A_{(0)}=0}$ for all three sets of modes in Eq.~(\ref{eq:RN_R_ob}), since by definition the inner horizon observer only sees positive frequency waves there. For the interior observer modes ${{}^\text{int}R^\pm_\text{ob}(z)}$, ${B_{(0)}=\exp(-i\omega r_-)/r_-}$, while the exterior observer modes ${{}^\text{ext}R^+_\text{ob}(z)}$ are only defined for ${z\geq1}$ and must be treated separately.

Unfortunately, analytic asymptotic forms for the modes of Eq.~(\ref{eq:Z_z0}) are not known at the spacetime's two other singular points. An explicit solution to the central two-point connection problem for confluent Heun functions is still outstanding and is directly related to the inverse of Hilbert's 21st problem; currently, analytic forms of the monodromy matrices have only been found for the reduced confluent Heun equation with ${\varepsilon=0}$ \cite{kaz06}.

Thus, we proceed by defining a new set of local Heun modes at each singular point and numerically matching their coefficients via the algorithm set forth in Ref.~\cite{mot18}.

At the event horizon (${z=1}$), a set of regular, linearly independent solutions to Eq.~(\ref{eq:heunc}) that are convergent everywhere except at the singular points ${z=0,\infty}$ can be written as:
\begin{subequations}\label{eq:Z_z1}
\begin{align}
    Z_{(1)}&(z)=A_{(1)}Z^A_{(1)}(z)+B_{(1)}Z^B_{(1)}(z),\\
    Z^A_{(1)}&(z)=\text{HeunC}\left(q-\alpha,-\alpha,\delta,\gamma,-\varepsilon;1-z\right),\\
    Z^B_{(1)}&(z)=(1-z)^{1-\delta}\nonumber\\
    \times&\text{HeunC}\left(q'-\alpha',-\alpha',2-\delta,\gamma,-\varepsilon;1-z\right),
\end{align}
\end{subequations}
with arbitrary complex coefficients $A_{(1)}$ and $B_{(1)}$, and with the definitions
\begin{subequations}
\begin{align}
    q'&=q-\gamma(1-\delta),\\
    \alpha'&=\alpha+\varepsilon(1-\delta).
\end{align}
\end{subequations}

Asymptotically, the modes of Eq.~(\ref{eq:Z_z1}) at the event horizon (${z=1}$) reduce to
\begin{equation}\label{eq:R_z1}
    R(z)\to\text{e}^{\frac{\varepsilon}{2}}|z-1|^{-\frac{1-\delta}{2}}\left(A_{(1)}+B_{(1)}(1-z)^{1-\delta}\right),
\end{equation}
which leads to the matching
\begin{subequations}\label{eq:AB_z1}
\begin{align}
    {}^\text{ext}A_{(1)}^+&=0,\nonumber\\
    {}^\text{ext}B_{(1)}^+&=\mathcal{T}^+_\text{ext}\frac{\text{e}^{i\omega(2r_+-r_-)}}{r_+};\displaybreak[0]\\{}^\text{int}A_{(1)}^+&=\left(\mathcal{T}^+_\text{int}+\mathcal{R}^+_\text{int}\mathcal{R}^-_\text{ext}\right)\frac{\text{e}^{-i\omega r_-}}{r_+},\nonumber\\
    {}^\text{int}B_{(1)}^+&=\mathcal{R}^+_\text{int}\frac{\text{e}^{i\omega(2r_+-r_-)}}{r_+};\displaybreak[0]\\
    {}^\text{int}A_{(1)}^-&=\mathcal{T}^-_\text{int}\frac{\text{e}^{-i\omega r_-}}{r_+},\nonumber\\
    {}^\text{int}B_{(1)}^-&=\left(\mathcal{R}^-_\text{int}+\mathcal{T}^-_\text{int}\mathcal{R}^-_\text{ext}\right)\frac{\text{e}^{i\omega(2r_+-r_-)}}{r_+};
\end{align}
\end{subequations}
for each respective set of modes; i.e., the coefficients from Eq.~(\ref{eq:Z_z1}) for ${{}^\text{ext,int}R^\pm_\text{ob}}$ are labeled ${{}^\text{ext,int}A^\pm_{(1)}}$ and ${{}^\text{ext,int}B^\pm_{(1)}}$. Eqs.~(\ref{eq:AB_z1}) are strictly only valid for ${z<1}$; for the exterior (${z>1}$), an additional factor of ${\exp[2\pi\omega r_+^2/(r_+-r_-)]}$ must be included in the right-hand side of the equations for each of the $B$ coefficients to account for the lack of absolute values in the trailing factor of Eq.~(\ref{eq:R_z1}).

At some point $z_*$ in the interior (we take ${z_*=0.5}$ for simplicity), both Eqs.~(\ref{eq:Z_z0}) and (\ref{eq:Z_z1}) provide regular solutions to the wave Eq.~(\ref{eq:heunc}). One can convert between them with the respective linear systems
\begin{subequations}
\begin{align}
    &Z^{A,B}_{(1)}(z_*)=C_A^{A,B}Z^A_{(0)}(z_*)+C_B^{A,B}Z^B_{(0)}(z_*),\\
    &(Z^{A,B}_{(1)})'(z)\big|_{z_*}\nonumber\\
    &=C_A^{A,B}(Z^A_{(0)})'(z)\big|_{z_*}+C_B^{A,B}(Z^B_{(0)})'(z)\big|_{z_*}.
\end{align}
\end{subequations}
The functions $Z^A_{(0)}(z_*)$, $Z^B_{(0)}(z_*)$, $Z^A_{(1)}(z_*)$ and $Z^B_{(1)}(z_*)$ can be computed numerically, and therefore the constants $C^A_A$, $C^B_A$, $C^A_B$, and $C^B_B$ can also be computed. Once these constants are known, the total eigenmodes $Z_{(0)}(z)$ and $Z_{(1)}(z)$ can be matched to solve for each of the backscattering coefficients:
\begin{subequations}
\begin{align}
    A_{(0)}&=A_{(1)}C_A^A+B_{(1)}C_A^B,\\
    B_{(0)}&=A_{(1)}C_B^A+B_{(1)}C_B^B.
\end{align}
\end{subequations}
Once the backscattering coefficients connecting ${z=0}$ to ${z=1}$ are known, a similar process will yield the coefficients connecting ${z=1}$ to ${z=\infty}$. As $z$ approaches infinity, the confluent Heun solutions to Eq.~(\ref{eq:heunc}) asymptotically (in a sector) take the form
\begin{subequations}\label{eq:Z_zinf}
\begin{align}
    &Z_{(\infty)}(z)=A_{(\infty)}Z^A_{(\infty)}(z)+B_{(\infty)}Z^B_{(\infty)}(z),\\
    &Z^A_{(\infty)}(z)=z^{-\frac{\alpha}{\varepsilon}},\\
    &Z^B_{(\infty)}(z)=\text{e}^{-\varepsilon z}z^{\frac{\alpha}{\varepsilon}-\gamma-\delta},
\end{align}
\end{subequations}
with arbitrary complex coefficients $A_{(\infty)}$ and $B_{(\infty)}$. Comparison with the asymptotic forms of Eq.~(\ref{eq:RN_R_ob}) reveals the following matched values for these coefficients:
\begin{subequations}
\begin{align}
    {}^\text{ext}A_{(\infty)}^+&=\mathcal{R}^+_\text{ext}\frac{\text{e}^{-i\omega r_-}}{r_+-r_-},\nonumber\\
    {}^\text{ext}B_{(\infty)}^+&=\frac{\text{e}^{i\omega r_-}}{r_+-r_-};\displaybreak[0]\\
    {}^\text{int}A_{(\infty)}^+&=0,\nonumber\\
    {}^\text{int}B_{(\infty)}^+&=\mathcal{R}^+_\text{int}\mathcal{T}^-_\text{ext}\frac{\text{e}^{i\omega r_-}}{r_+-r_-};\displaybreak[0]\\
    {}^\text{int}A_{(\infty)}^-&=\mathcal{T}^-_\text{int}\mathcal{T}^-_\text{ext}\frac{\text{e}^{-i\omega r_-}}{r_+-r_-},\nonumber\\
    {}^\text{int}B_{(\infty)}^-&=0,
\end{align}
\end{subequations}
where the coefficient notation is the same as in Eq.~(\ref{eq:AB_z1}).

For some sufficiently large radial coordinate ${z=z^*}$ (we find heuristically that ${z^*=18/(\omega\sqrt{1-Q^2})}$ is more than sufficient to ensure convergence at machine-level precision), both Eqs.~(\ref{eq:Z_z1}) and (\ref{eq:Z_zinf}) satisfy the wave Eq.~(\ref{eq:heunc}), and so the two sets of solutions can be matched. One has the system
\begin{subequations}
\begin{align}
    &Z^{A,B}_{(1)}(z^*)=D_A^{A,B}Z^A_{(\infty)}(z^*)+D_B^{A,B}Z^B_{(\infty)}(z^*),\\
    &(Z^{A,B}_{(1)})'(z)\big|_{z^*}\nonumber\\
    &=D_A^{A,B}(Z^A_{(\infty)})'(z)\big|_{z^*}+D_B^{A,B}(Z^B_{(\infty)})'(z)\big|_{z^*}
\end{align}
\end{subequations}
to solve for the constants $D^A_A$, $D^B_A$, $D^A_B$, and $D^B_B$, which can then be used to solve for the backscattering coefficients with the system 
\begin{subequations}
\begin{align}
    A_{(\infty)}&=A_{(1)}D_A^A+B_{(1)}D_A^B,\\
    B_{(\infty)}&=A_{(1)}D_B^A+B_{(1)}D_B^B.
\end{align}
\end{subequations}
Altogether, the relevant backscattering coefficients can be written as follows (note that multiple variations to the below equations are possible based on implicit relations between and among the $C$ and $D$ coefficients):
\begin{subequations}
\begin{align}
    \mathcal{T}^+_\text{ext}&=\frac{r_+}{r_+-r_-}\frac{1}{D_B^B}\ \text{e}^{-2i\omega(r_+-r_-)-\pi\omega/\varkappa_+},\\
    \mathcal{T}^-_\text{ext}&=\frac{r_+-r_-}{r_+}\frac{\widetilde{D}}{D_B^B},\displaybreak[0]\\
    \mathcal{R}^+_\text{ext}&=\frac{D_A^B}{D_B^B}\ \text{e}^{2i\omega r_-},\\
    \mathcal{R}^-_\text{ext}&=-\frac{D_B^A}{D_B^B}\ \text{e}^{-2i\omega r_+-\pi\omega/\varkappa_+},\displaybreak[0]\\
    \mathcal{T}^+_\text{int}&=-\frac{r_+}{r_-}\frac{C_A^BD_B^B-C_A^AD_B^A\ \text{e}^{-4i\omega r_+-\pi\omega/\varkappa_+}}{\widetilde{C}D_B^B},\\
    \mathcal{T}^-_\text{int}&=-\frac{r_+}{r_-}\frac{C_A^B}{\widetilde{C}},\displaybreak[0]\\
    \mathcal{R}^+_\text{int}&=\frac{r_+}{r_-}\frac{C_A^A}{\widetilde{C}}\ \text{e}^{-2i\omega r_+},\\
    \mathcal{R}^-_\text{int}&=\frac{r_+}{r_-}\frac{C_A^AD_B^B-C_A^BD_B^A}{\widetilde{C}D_B^B}\ \text{e}^{-2i\omega r_+},
\end{align}
\end{subequations}
where
\begin{align}
    \widetilde{C}&\equiv C_A^AC_B^B-C_A^BC_B^A,\\
    \widetilde{D}&\equiv D_A^AD_B^B-D_A^BD_B^A.
\end{align}
The resulting numerical values of the backscattering coefficients are used to calculate the Hawking spectra of Fig.~(\ref{fig:spectra}).

\bibliography{apsbib}

\end{document}